\documentclass[aps,article,twocolumn,superscriptaddress,showpacs]{revtex4}
\usepackage{epsfig}
\usepackage{natbib}
\usepackage{amsmath}
\usepackage{times}
\usepackage{psfrag}
\usepackage{subfigure}
\usepackage{graphicx}
\usepackage{amsthm}
\usepackage{subfigure}
\usepackage{epstopdf}
\usepackage{times}

\newcommand{\hg}{ }

\begin{document}

\title{Noise and disorder effects in a series of birhythmic Josephson junctions
 coupled to a  resonator}

\author{O. V. Pountougnigni}
\affiliation{Laboratory of Mechanics and Materials, Department of Physics, Faculty of
Science,
University of Yaound\'e I, Box 812, Yaound\'e, Cameroon.}
\author{R. Yamapi}
\email[Author to whom correspondence should be addressed. Electronic mail : ] {ryamapi@yahoo.fr}
\affiliation{Fundamental Physics Laboratory, Physics of Complex System group, Department of Physics, Faculty of
 Science, University of Douala, Box 24 157 Douala, Cameroon.}

\author{G. Filatrella}
 \affiliation{Department  of Sciences and Technologies,
 and Salerno unit of CNSIM, University of Sannio, Via Port'Arsa 11,
I-82100 Benevento, Italy.}

\author{C. Tchawoua}
\affiliation{Laboratory of Mechanics and Materials, Department of Physics, Faculty of Science,
 University of Yaound\'e I, Box 812, Yaound\'e, Cameroon.}

\begin{abstract}

\hg {This paper deal the effects of
uncorrelated white noise, in a serie of Josephson Junctions coupled to a linear $RLC$ resonator.
The junction are hysteretic, and hence can be considered birhythmic,
that is capable to oscillate at different frequencies for the same set of parameters.
Both  Josephson Junctions with identical and
 disordered parameters are considered.
With the uniform parameters, the array behaves similarly to
single Josephson junctions, also in the presence of noise.
The magnitude of the effective energy that characterizes the response to noise becomes smaller as the number of elements of the array increases, making the resonator less stable.
Disorder in the parameters drastically changes the physics of the array.
The disordered array of Josephson junctions misses the birhythmicity properties for large values of the variance of the disorder parameter.
Nevertheless, the system remains birhythmic for low values of the disorder parameter.
Finally, disorder makes it difficult to locate the separatrix, hinting to a more complex structure of the effective energy landscape. }\\
\textbf{\emph{Keywords}}:\emph{Josephson junctions; Mean First Passage time, Synchronization }

\end{abstract}
\pacs{05.45.Xt, 05.40.Ca, 85.25.Cp}
\date{\today}
\maketitle

\section{Introduction }
\label{introduction}
{\hg
Hysteretic (large capacitance) Josephson Junction (JJ) can be considered  a birhythmic
system,  they can produce oscillations at two distinct periods \cite{Barone82,Jain84}.
Birhythmicity is encountered in some biochemical \cite{Decoroly82,Morita89,Haberichter01,Sosnovtseva02,Abou11} and non linear electronic systems \cite{Kadji07,Zakharova10,Yamapi10,Ghosh11,Yamapi12,Yue12}.
In JJ physics, it is encountered in arrays coupled through an external circuit that possesses resonances \cite{Likharev86}.
In this condition, the array can either oscillate at two frequencies, the one induced by the external resonance that locks together the JJ, or at the spontaneous frequency of JJ where the elements oscillate incoherently and are unable to load the external circuit.
{ Synchronization of JJ oscillators is also of practical importance for applications in which the power of a single JJ does not suffice.}
In fact, the power of $N$ coherently working junctions can increase as $N^2$, and the linewidth can also decrease as $1/N$ \cite{Cawthorne99,Barbara99,Vasilic01} .
The most frequent geometric configuration is probably a two-dimensional array (a combination of parallel and series elements) \cite{Barbara99,Filatrella01} and   one-dimensional array (parallel or series array) \cite{Filatrella03}.
However, synchronization is not easily obtained, for even with the same bias current the different junctions oscillate at different voltages because of the differences in the fabrication parameters \cite{Tolpygo15,Tolpygo17}.
Synchronization of JJ can be induced by an appropriated external coupling circuit \cite{Hadley88} that can be realized in a variety of manners \cite{Jain84}.
 For example in \cite{Filatrella92} each junction is coupled to the resonator and  the junctions are coupled to each other through the capacitance.
Possible coupling mechanisms are $RLC$ circuits \cite{Tachiki11,Welp13}, or $LC$ circuits \cite{Shukrinov17}, that allow the possibility of parametric resonance \cite{Shukrinov12}.
Thus, it has been possible to derive the properties of the synchronized and unsynchronized states, to characterize the conditions that favor synchronization, especially the coupling circuit configuration and the bias point \cite{Shukrinov12,Shukrinov17}, that are the parameters most easily tuned in experiments.

Having established the configuration, the global stability properties are of crucial importance to  determine the region of parameters where realistic arrays, that is disordered and noisy elements, could  possibly work.
The problem of synchronization of disordered oscillators, including JJ arrays, is often dealt with the Kuramoto model and its variations \cite{Rodrigues16}.
The Kuramoto model, that perhaps should be named a {\it framework} for the numerous variants, has been extended in many directions, for instance to incorporate finite size and interplay between different types of noise and couplings \cite{Rodrigues16}.
%

Thus the Kuramoto model allows to retrieve the properties of the synchronized states.
 However, a different problem, similar to Kramer's escape, can arise: under the effect of noise, how often the system experiences large excursions, large enough to move from the synchronous to the unsynchronized state?
 These switches are of particular relevance in birhythmic systems, for the noiseless attractors are characterized by different frequencies and therefore a marked change in the frequency is associated to such large excursions.
The problem is pertinent to the application of the JJ to the voltage standard, for the frequency of the oscillations is connected, through the Josephson relation, to the junction voltage.
Thus, even very rare escapes are of practical relevance to reach the extreme accuracy demanded by international standards.
The stability analysis of the noisy oscillations dynamical behavior can be summarized by means of a quasi- or  pseudo-potential \cite{Graham86,Bouchet16} when the system does not possesses a potential, that is the force cannot be derived as the gradient of function.
The quasipotential is the effective energy barrier that governs the low noise escapes from the metastable states  \cite{Graham86,Bouchet16}; the concept has been proved useful in connection with JJ based voltage standard \cite{Kautz88,Kautz96}.

Investigated these large random excursions under the effect of noise arises the conceptual difficulty of the very definition of {\it attractor} and {\it limit cycle}.
In fact under the effect of noise the system is not confined, even asymptotically, in a well defined limit cycle, for the noise disturbance that affects the dynamics.
{ Put it in other words, the noiseless limit cycles characterized by a well defined frequency, under the effect of noise become blurred and ill defined.
The abstract definition does not suffer of this problem, as it entails the condition for which the attractor has been abandoned (i.e., the time necessary to reach the separatrix) in the limit of negligible noise (that is, small enough that the concept of sepratrix retains its sense).
This  condition, that is well defined in a mathematical sense, poses a problem in the numerical evaluation, that of course are performed at finite noise value, for the evaluation of the quasi-potential requires to determine the average time necessary to leave an attractor, that is the mean first escape time ({MFPT}) to reach the separatrix.
In practical terms, the difficulty is circumvented controlling that the noise is low enough that the uncertainty due to the approximate definition of the separatrix  introduces a negligible error in the determination of the time necessary for a passage from a region of the  phase space to the other.

A further problem is to find such border that separates the two attractors, even in the limit of negligible noise, for a system of many oscillators}, and hence of many degrees of freedom, the separatrix consists of a complicated hyper-surface in $N-1$ dimensions, if $N$ is the number of degrees of freedom.
To make the problem manageable, it has been proposed to approximate the position of the separatrix with the region where the slope of the mean first passage time changes \cite{Yamapi14}.
However, it remains an open problem, the extension of the method to
the case of many disordered JJs, that is the subject of the present work.

 The paper is organized as follows.
 The next Section describes a serie of  underdamped disordered {Josephson Junction }; both connected to an $RLC$  resonator and subject to external bias and noise.
Section \ref{deterministic} focuses on the properties of the attractors in a birhythmic {Josephson Junction } array, and especially in the method to locate the effective separatrix, which is, as mentioned, the  essential information to construct the activation energy barrier.
The effective potential that determines the global stability of the system is studied.
 In Sect. \ref{disorder}, the effects of disordered parameters are analysed on the birhythmic properties.
Section \ref{conclusions} closes the paper with conclusions.

\section{Model of a serie of Josephson Junctions coupled to a resonator}
\label{model}
This Section describes the basic model in Sect. \ref{modelJJ} and the refinements to include noise and disorder in Sect. \ref{modeldisorder} . The numerical method is described in the last part, Sect. \ref{algorithm}.
}
    \subsection{Model of Josephson junctions dynamics}
\label{modelJJ}

\begin{figure}
\begin{center}
\includegraphics[height=4.2cm,width=7.8cm]{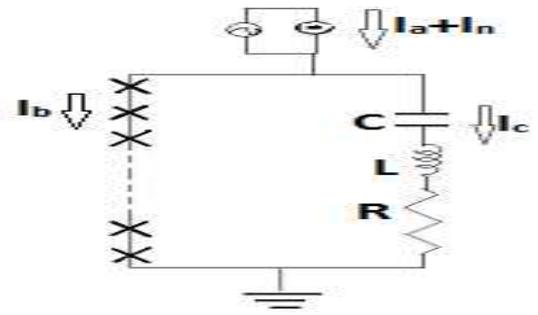}
\caption{\it {\hg (a) Scheme of a series of birhythmic Josephson junctions coupled to a linear $RLC$ resonator.
 (b) Circuit model, RCSJ, of a Josephson junction.}
The current bias is at room temperature, and it is supposed to supply an ideal current source $I_a$ and a noisy Gaussian distributed current $I_n$.
The junctions and the $RLC$ resonator are in the refrigerated box, and the associated Johnson noise is supposed negligible.
}
\label{circuit}
\end{center}
\end{figure}

\begin{figure}
\begin{center}
\includegraphics[scale = 0.33]{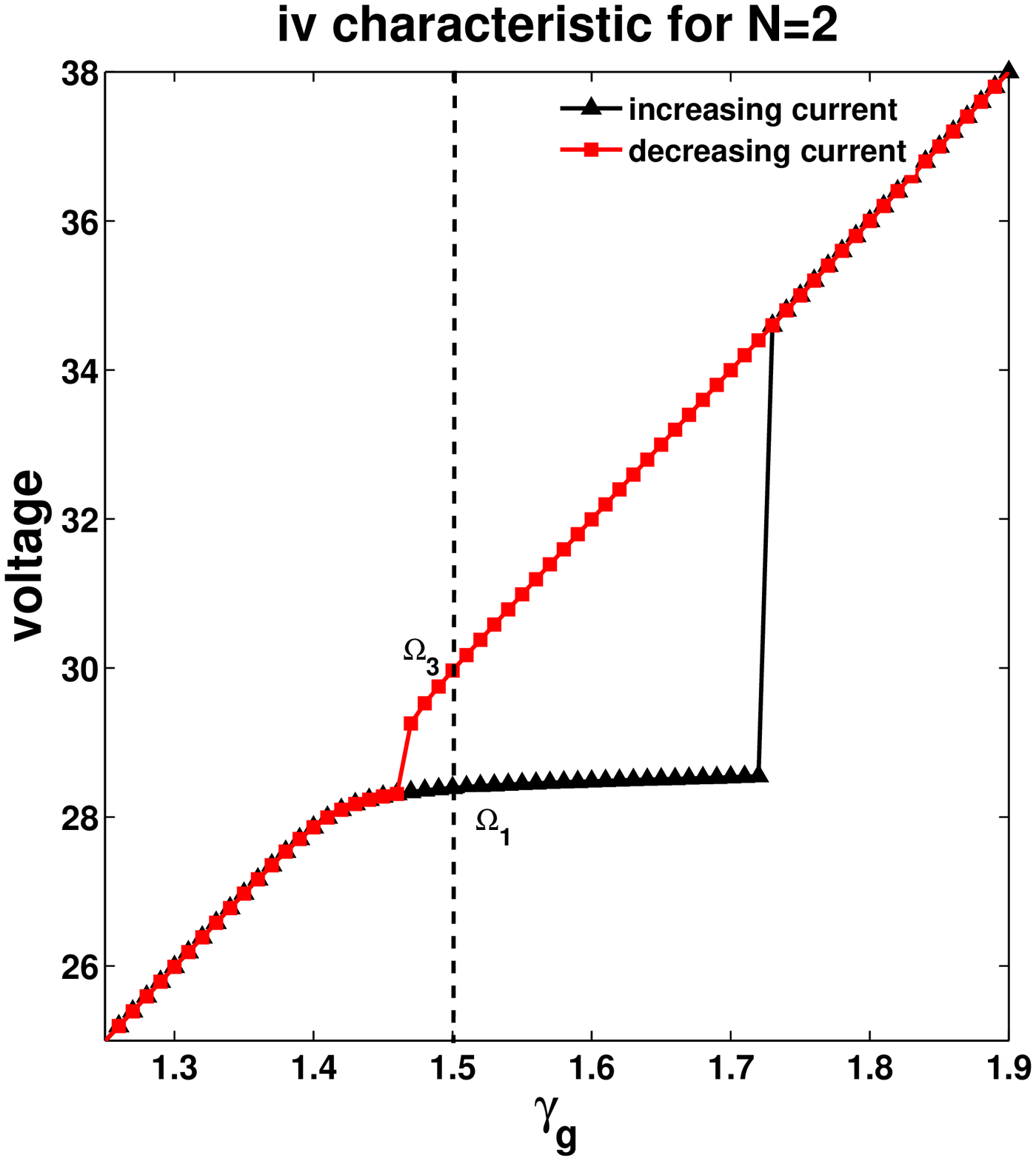}
\includegraphics[scale = 0.33]{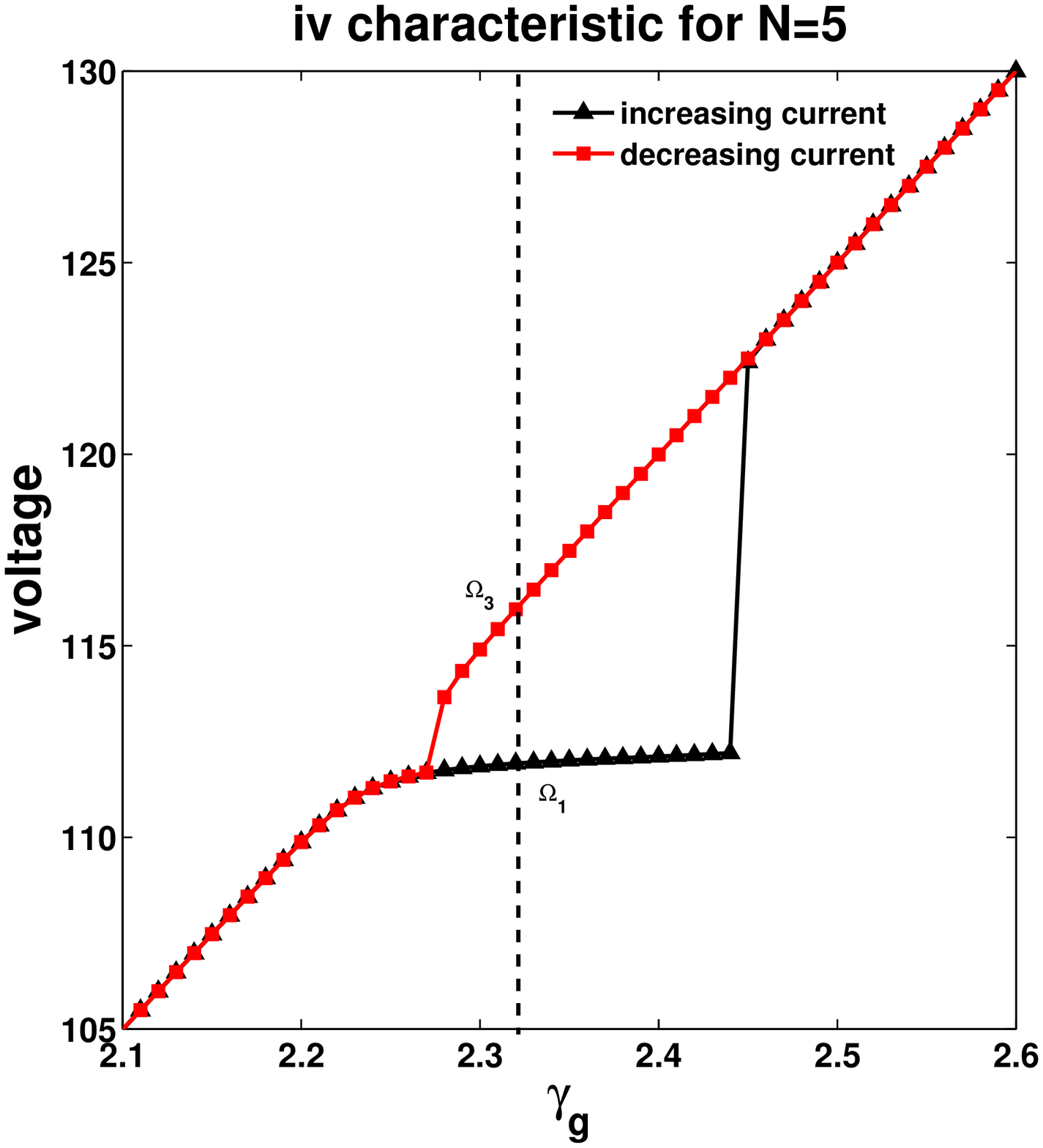}
\caption{\it Normalized IV  curves for both increasing and decreasing normalized bias current $\gamma_G$ for  the free noise model, $D=0$.
The parameters are $\beta_L=0.01$, $Q=200$, $\Omega=2.0$,
  $\alpha=0.1$; for $(i):N=2$ and for  $(ii):N=5.$}
\label{I-V}
\end{center}
\end{figure}

 Figure \ref{circuit} schematically describes the model: a serie
  of  underdamped JJ coupled to an $RLC$ resonator.
Both the JJs and the resonant circuit are supposed in the temperature controlled vessel,
 while the bias current is supplied by a device typically at room temperature.
 In this configuration, the noise from the bias supply dominates respect to the Johnson noise from the resistors $R_J$ and $R$.
Alternatively, one could add a random term for each resistor, as done for instance in Ref.
\cite{lin11}.
However, the noise is but a tool. Our goal is to determine the pseudo-energy; the principle of minimum energy \cite{Graham86,Kautz88} assures that the contributions from the minimal trajectory determines the height of the trapping potential, and therefore one does not expect substantial changes with a different noise source.

The electrical model consists of the capacitor $C_J$, the resistor $R_J$, and the ideal Josephson element, connected in series.
The nonlinear relation between the current and the gauge invariant phase difference
 $\phi^i =\phi_1-\phi_2$ across two superconductors:
\begin{eqnarray}
\label{eqJcurrent}
I_J^i=I_0\sin \phi^i,
\end{eqnarray}
together with the Josephson voltage relationship
\begin{eqnarray}
\label{eqJvolt}
V_J^i=\frac{\hbar}{2e}\frac{d\phi^i}{dt},
\end{eqnarray}
 determines that a JJ is an active oscillator that converts a $dc$
 current into an $ac$ drive for the $RLC$ resonator.
 {\hg As mentioned in the introduction, $RLC$ coupling in parallel to the array is but a model for the external circuit that embeds the JJs.
 In fact, at the microwave frequency the lumped elements is an approximation of the distributed elements.
 In this context, the choice of series and parallel depends upon the relative magnitude of the impedance of the JJ; for low damping the external circuit ``looks parallel'', while for high damping the external circuit ``looks series'' \cite{Likharev86}.
In this work we concentrate on the particular setting of $RLC$ series oscillator.
We emphasize that this is but an option that we employ for brevity to illustrate the method.
This coupling is particularly convenient for it shows a single branch connected with the $RLC$ resonance -- for the present purposes it is advantageous to have only two regions in the phase space with two corresponding frequencies.
Other possibilities, as $LC$  coupling in stacks  are useful for the description of more complicated structures, e.g., of traveling waves in stacks of JJs \cite{Shukrinov12}.}

 To derive the equations governing the system, we indicate with $I_C$ the current flowing through the $RLC$ circuit and with $\tilde q$ the charge on the capacitor.
 The JJ array and the resonator are both biased by a current generator  $I_G$ affected by a noise current $I_n$ that split in the current $I_b$ through the \textbf{JJ} element and the above mentioned current $I_C$ through the $RLC$.
 If we indicate with $I_{R_J}$ the current through the JJ resistor and with $I_{C_J}$ the
current through the junction capacitance, one obtains the current balance:
\begin{equation}
\label{currbalance}
I_b=I_a+I_n-I_C
\Longrightarrow I_J^i+I_{R_J}+I_{C_J}=I_a+I_n-I_C.
\end{equation}
 The Kirchhoff law for the loop voltage
\begin{eqnarray}
\label{voltRLC}
\sum_{i=1}^N V_J^i=V_C+V_{R}+V_L
\end{eqnarray}
completes the model, that is thus described by two second order coupled differential
equations:
\begin{eqnarray}
\label{eqJJ_RLC}
\left\{\begin{array}{l}
\frac{C_J\hbar}{2e}\frac{d^2\phi^i}{dt^2}+\frac{\hbar}{R_J 2e}\frac{d\phi^i}{dt}+I_0\sin\phi^i+\frac{d\tilde q}{dt}
= I_a+I_n\\
\frac{d^2\tilde q}{dt^2}+\frac{R}{L}\frac{d\tilde q}{dt}+
\frac{1}{LC}\tilde q
-  \frac{\hbar}{2eL}  \sum_{i=1}^N\frac{d\phi^i}{dt}
= 0.
\end{array}\right.
\end{eqnarray}

Introducing the Josephson frequency $\omega_J=\sqrt{2eI_0 / C_J\hbar}$, Eqs.(\ref{eqJJ_RLC}) can be cast, with normalized time $\tau=\omega_J t$ and charge $q = \omega_J \tilde q /I_0$, as follows:
\begin{eqnarray}
\label{eqJJ_RLC_norm}
\left\{\begin{array}{l}
\frac{d^2\phi^i}{d\tau^2}+\alpha\frac{d\phi^i}{d\tau}+\sin\phi^i+\frac{d q}{d\tau}=\gamma_G+\zeta^i ({\tau}),\qquad i=1,2,...,N\\
\frac{d^2 q}{d\tau^2}+\frac{1}{Q}\frac{d q}{d\tau}+\Omega^2 q  -
\frac{1}{\beta_L} \sum_{i=1}^N\frac{d\phi^i}{d\tau}=0,
\end{array}
\right.
\end{eqnarray}
where the parameters are defined as
\begin{equation}
\label{parameters}
Q=\frac{L\omega_J}{R},
\quad \alpha=\frac{1}{C_J\omega_J}\sqrt{\frac{1}{R_J}},
\quad \beta_L=\frac{\omega_J^2C_J}{L},
\quad \gamma_G=\frac{I_a}{I_0},
\quad \Omega = \frac{1}{\omega_J\sqrt{LC}}.
\end{equation}

{\hg The described coupling circuit is thus formed by two branches, a JJ and an $RLC$ resonator in a loop, strongly interacting each other, as the coupling is large, $1\slash \beta_L = 100$.
The two branches are thus not completely independent, the resonance of the circuit, for instance, is away from  the $RLC$ resonance $\Omega$, as shall be discussed in connection with Figs. \ref{I-V} in the next Sect. \ref{attractors}.
}

\begin{figure}
\begin{center}
\includegraphics[scale = 0.33]{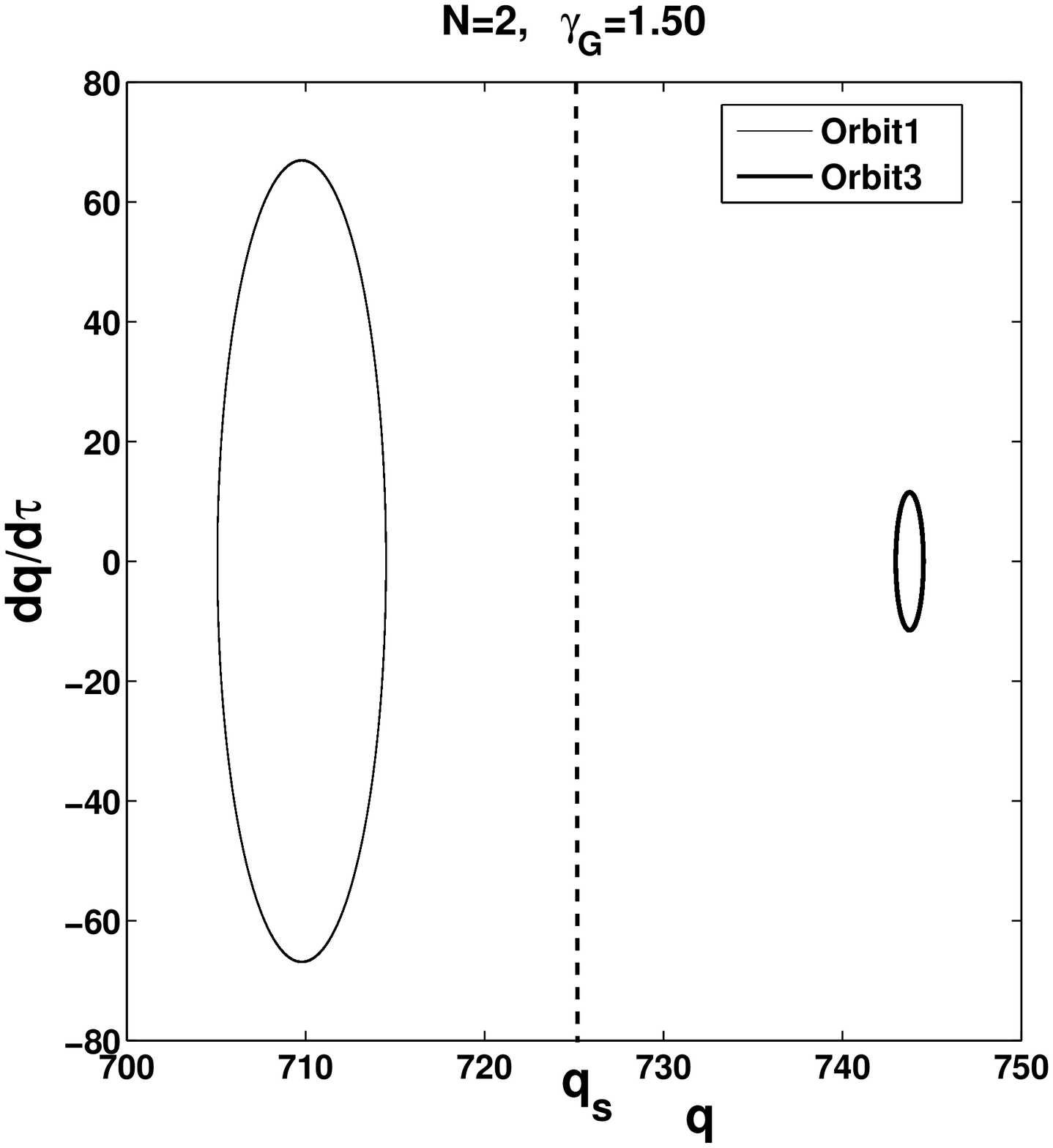}
\includegraphics[scale = 0.33]{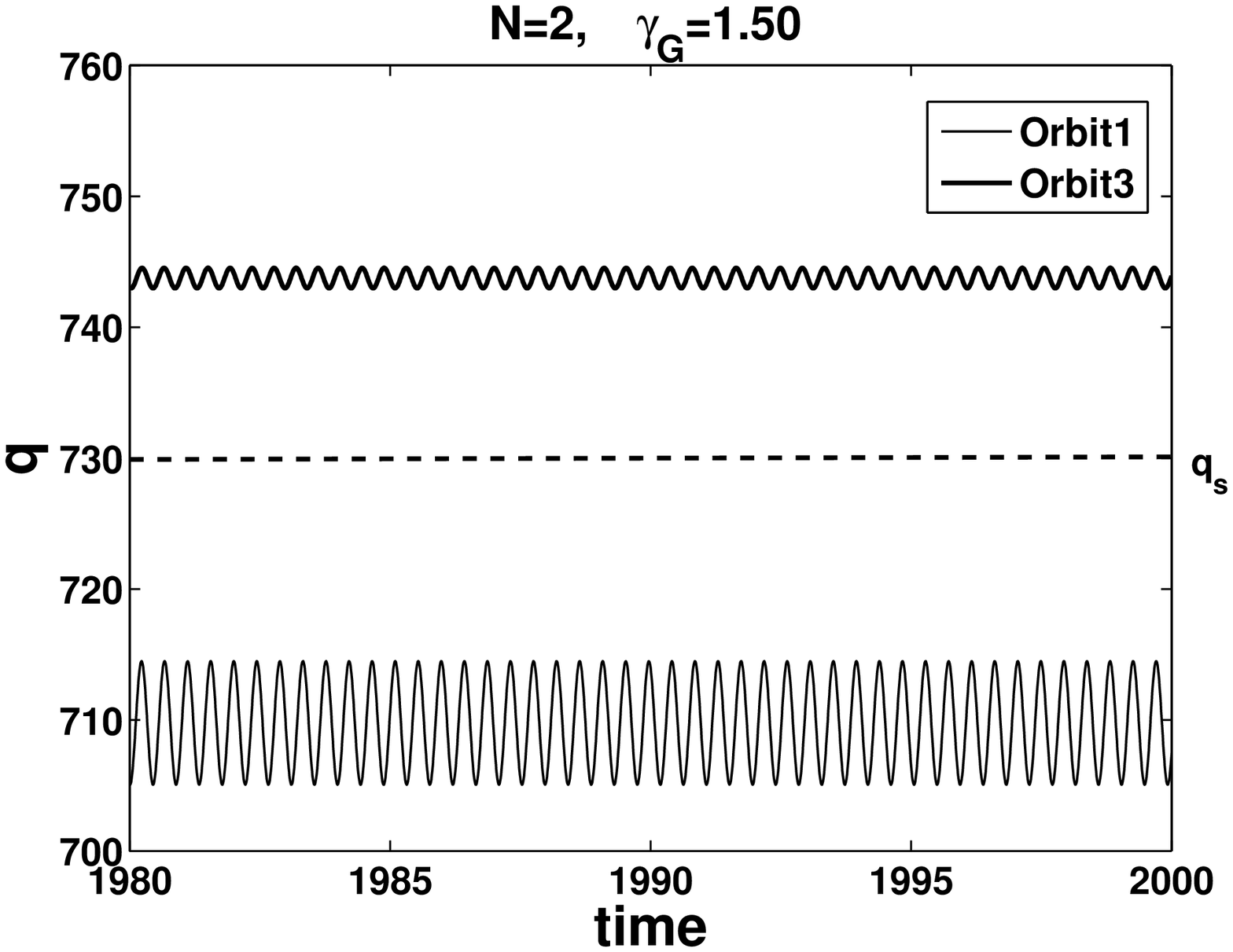}
\includegraphics[scale = 0.33]{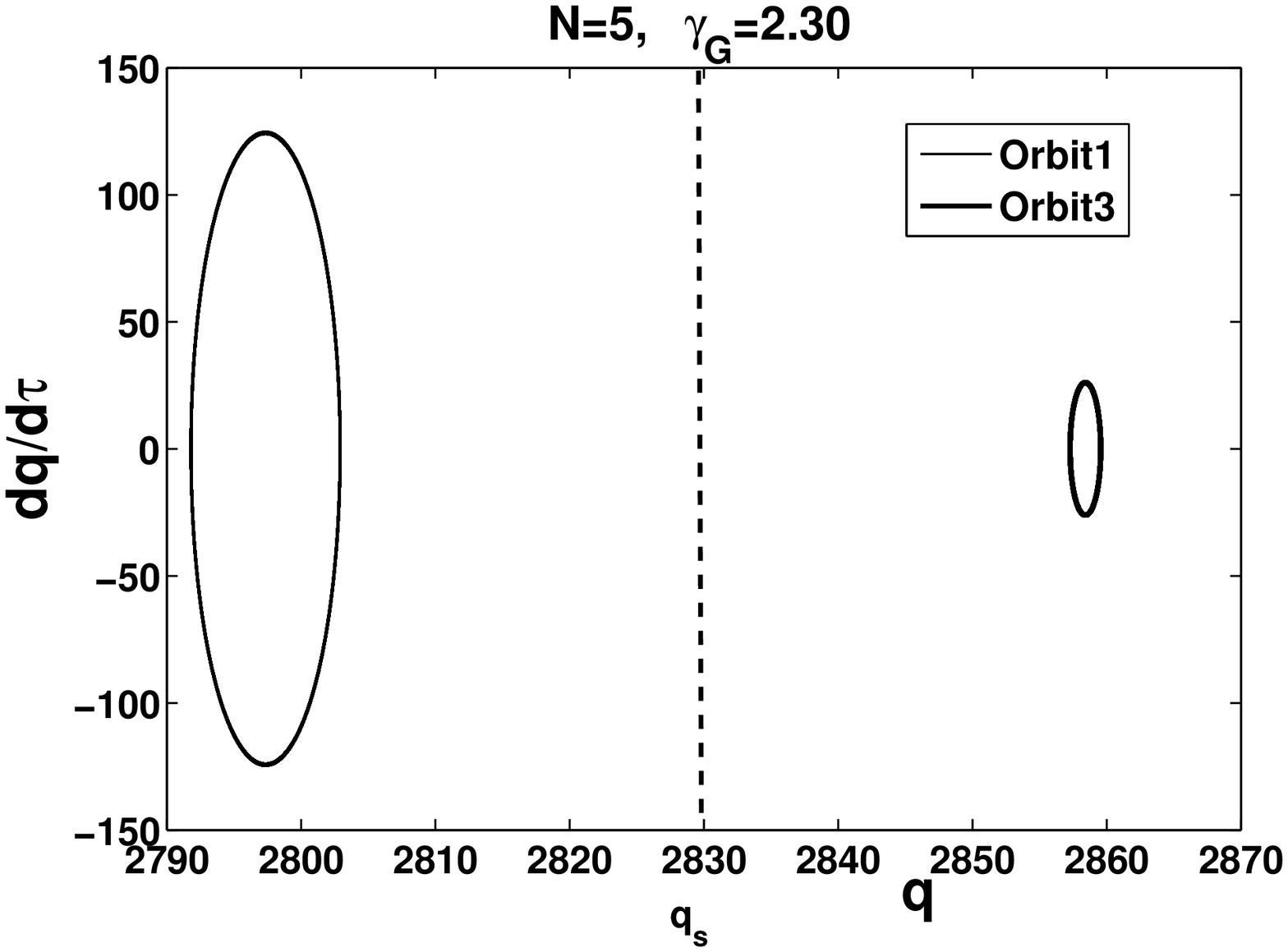}
\includegraphics[scale = 0.33]{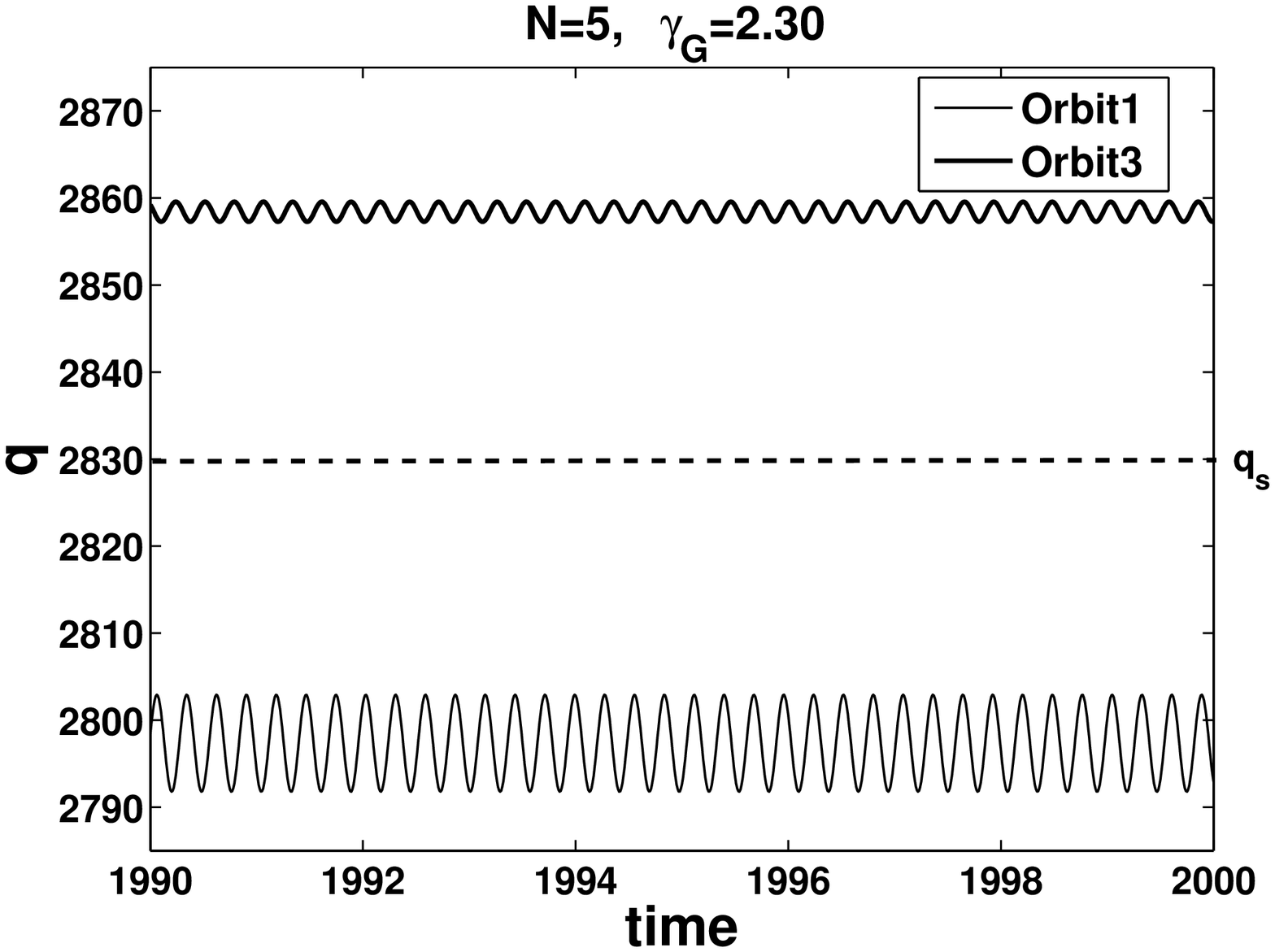}
\caption{\it Projection of the phase space in the $q-dq/d\tau$ plane and time evolution for the free noise coupled JJ model, $D=0$.
Here $q_s$ indicates the hypothesized position of the separatrix.
The parameters of simulations are:
 $\beta_L=0.01$, $Q=200$, $\Omega=2.0$, and  $\alpha=0.1$.
 For (i) and (ii):
$N=2$, $\gamma_G=1.50$ ; for (iii) and (iv) : $N=5$, $\gamma_G=2.30$.}
\label{attract}
\end{center}
\end{figure}

The statistical features of the noisy term $\zeta^i$ are determined by:
\begin{eqnarray}
\label{correl}
<\zeta^i(\tau)> &=& 0, \nonumber \\
<\zeta^i(\tau);\zeta^i(\tau')> &=& 4 D \delta(\tau-\tau').
\end{eqnarray}
where  $D=k_B T \omega_J/\left(R I_0^2\right)$ is the normalized noise intensity ($k_B$ is the Boltzman constant and $T$ the absolute temperature).
If the Johnson noise associated to the resistances of each {JJ} are sizable, one has separated contributions $\zeta^1, \zeta^2 ...  \zeta^N $ for the junctions.
Also, Eq.(\ref{correl}) is changed, for the normalized noise intensity $D$ is given by the resistance $R_J$ of the junctions.
Instead, if the noise associated to the superconducting JJs are negligible respect to the resonator, then the noise comes from the resonator resistance $R$ and it is common to all the junctions.
Finally, if the noise is due to an external source, it does not obey to the fluctuation-dissipation theorem and it is therefore independent of the resistance.

\subsection{Model of disorder}
\label{modeldisorder}

In the fabrication process of several {JJ}, as in Fig. \ref{circuit},
fabrication tolerances prevent the junctions to be identical.
More precisely, as the barrier average thickness changes from junction to junction, the critical current changes most, as it depends exponentially on the junction barrier \cite{Barone82}.
Moreover, the resistance changes, for the product $R_J I_0$ is proportional to the superconducting gap.
To include these observations in the mathematical formulation of Sect. \ref{model}, one can assume that the resistance and the critical current depend upon the junction index $i$:
$R_J \rightarrow R_J^i$ and $I_0 \rightarrow I_0^i$, and therefore
\begin{equation}
\label{gap}
R^i_J I^ì_0 = \frac{\Delta}{e}.
\end{equation}
{\hg
The distribution of the critical currents of the JJ is here assumed uniform in an interval $\pm \varepsilon I_0$ around the average value, that we keep indicating with $I_0$.
The choice of the uniform distribution arises from two considerations.
The actual distribution of the parameters of junctions’ arrays, is close to Gaussian \cite{Tolpygo15,Tolpygo17}.
To avoid negative unphysical values of the critical current, the random variable $\varepsilon$ is assumed uniform in an interval of amplitude $2\sigma$:
}
\begin{equation}
\label{epsdistribution}
\varepsilon \in \left[-\sigma; + \sigma \right],
\end{equation}
and consequently the distribution of the critical currents of the $i^{th}$ junction reads
\begin{equation}
\label{I0distribution}
I^i_0 = I_0 \left( 1+\varepsilon^i\right).
\end{equation}
As a consequence of the relation (\ref{gap}), the resistance reads
\begin{equation}
\label{RJdistribution}
R^i_J = \frac{\Delta}{e} \frac{1}{I_0 \left( 1+\varepsilon^i\right)}.
\end{equation}

The normalized parameters (\ref{parameters}) are unchanged, but for the dissipation $\alpha$ that reads:
\begin{equation}
\label{alphadistribution}
\alpha^i  = \frac{1}{C_J\omega_J} \sqrt{\frac{1}{R_J^i} } =
\frac{1}{C_J\omega_J} \sqrt{\frac{e I_0 \left( 1+\varepsilon^i\right) }{\Delta} } =
\frac{1}{C_J\omega_J} \sqrt{\frac{ \left( 1+\varepsilon^i\right) }{R_J^i} }  =
\alpha  \left( 1+\varepsilon^i\right) .
\end{equation}
Finally, the equations for the disordered model Eqs.(\ref{eqJJ_RLC_norm}) read
\begin{eqnarray}
\label{eqJJ_RLC_disordered}
\left\{\begin{array}{l}
\frac{d^2\phi^i}{d\tau^2}+\alpha\left( 1 + \varepsilon^i\right) \frac{d\phi^i}{d\tau}+ \left( 1 + \varepsilon^i\right) \sin\phi^i+\frac{d q}{d\tau}=\gamma_G+\zeta^i ({\tau}),\qquad i=1,2,...,N\\
\frac{d^2 q}{d\tau^2}+\frac{1}{Q}\frac{d q}{d\tau}+\Omega^2 q
-  \frac{1}{\beta_L}  \sum_{i=1}^N \frac{d\phi^i}{d\tau}=0.
\end{array}
\right.
\end{eqnarray}

A stochastic model for the random parameters $I_0^i$ and $R_J^i$ is given by the assumptions (\ref{I0distribution}) and (\ref{RJdistribution}), thus, the random parameters depend on a single parameter $\varepsilon^i$ characterized by the uniform distribution (\ref{epsdistribution}), whose variance is $2\sigma^2 /3$.

As a further simplification, we here just assume that the parameters are uniformly distributed in the interval $[-\sigma; + \sigma]$ , that is:
\begin{equation}
\label{epsuniform}
\varepsilon^i =- \sigma + 2 \left( i-1\right) \frac{\sigma}{N-1}.
\end{equation}

As the Josephson fabrication tolerances are typically of the order of few percents \cite{Tolpygo15,Tolpygo17}, one can safely assume $\sigma = 0.1\% \div 10\%$.

\subsection{Algorithm for the numerical solution}
\label{algorithm}

Equations (\ref{correl},\ref{eqJJ_RLC_disordered}) are simulated with the Euler algorithm \cite{Fox88}.
Deterministic results have been obtained using the fourth order {Runge Kutta} algorithm \cite{lapidus}.
The stochastic results are averaged over as many realizations
as necessary to guarantee convergence in the statistical sense) within $5\%$.
The Gaussian white noise is generated using the {Box-Muller} algorithm \cite{Knuth69}  from two random numbers, $a$ and $b$, which are uniformly distributed on  the unit interval $[0,1]$.
Thus, for each step $\Delta \tau$, $\zeta^i_{n}$  is  obtained as follows:
\begin{eqnarray}
\label{random}
\nonumber a = \textrm{random  number},\quad b = \textrm{random  number}, \\
\zeta^i_{n} = \sqrt{-4D\Delta \tau \log(a)}\cos(2\pi b)
\end{eqnarray}

For some values of $N$  the $IV$ curves have been obtained by slowly increasing the current with a step $\Delta \gamma_{G }=0.01$,
and using the final state at the previous current step as the
initial step for the increased (or decreased) current biased
(see Fig.\ref{I-V}). At each current step a transient of about $1000$
normalized time is discarded. The averages are also calculated over the same time.
The time step $\Delta \tau$ is, through all simulations, $\Delta \tau=0.0001$ for the Euler algorithm and $\Delta \tau=0.01$ for the fourth order {Runge Kutta} algorithm.    .

\begin{figure}
\begin{center}
\includegraphics[scale = 0.35]{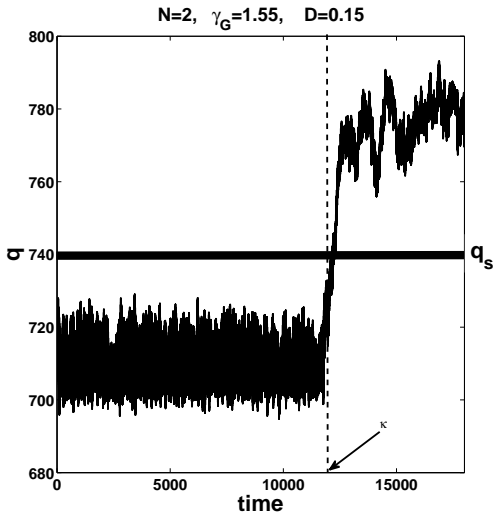}
\caption{\it Example of the switch from the attractor at frequency $\Omega_1$ (locked to the $RLC$ resonator) to the attractor at frequency $\Omega_3$ (unlocked state) under the influence of noise.
After the time $\kappa=12000$ normalized units the system crosses the estimated separatrix $q_s$.
The parameters are:
$\beta_L=0.01$, $Q=200$, $\Omega=2.0$,  $\alpha=0.1$, $\gamma_G=1.55$, $D=0.15$, $N=2$.}
\label{switch2}
\end{center}
\end{figure}

\begin{figure}
\begin{center}
\includegraphics[scale = 0.35]{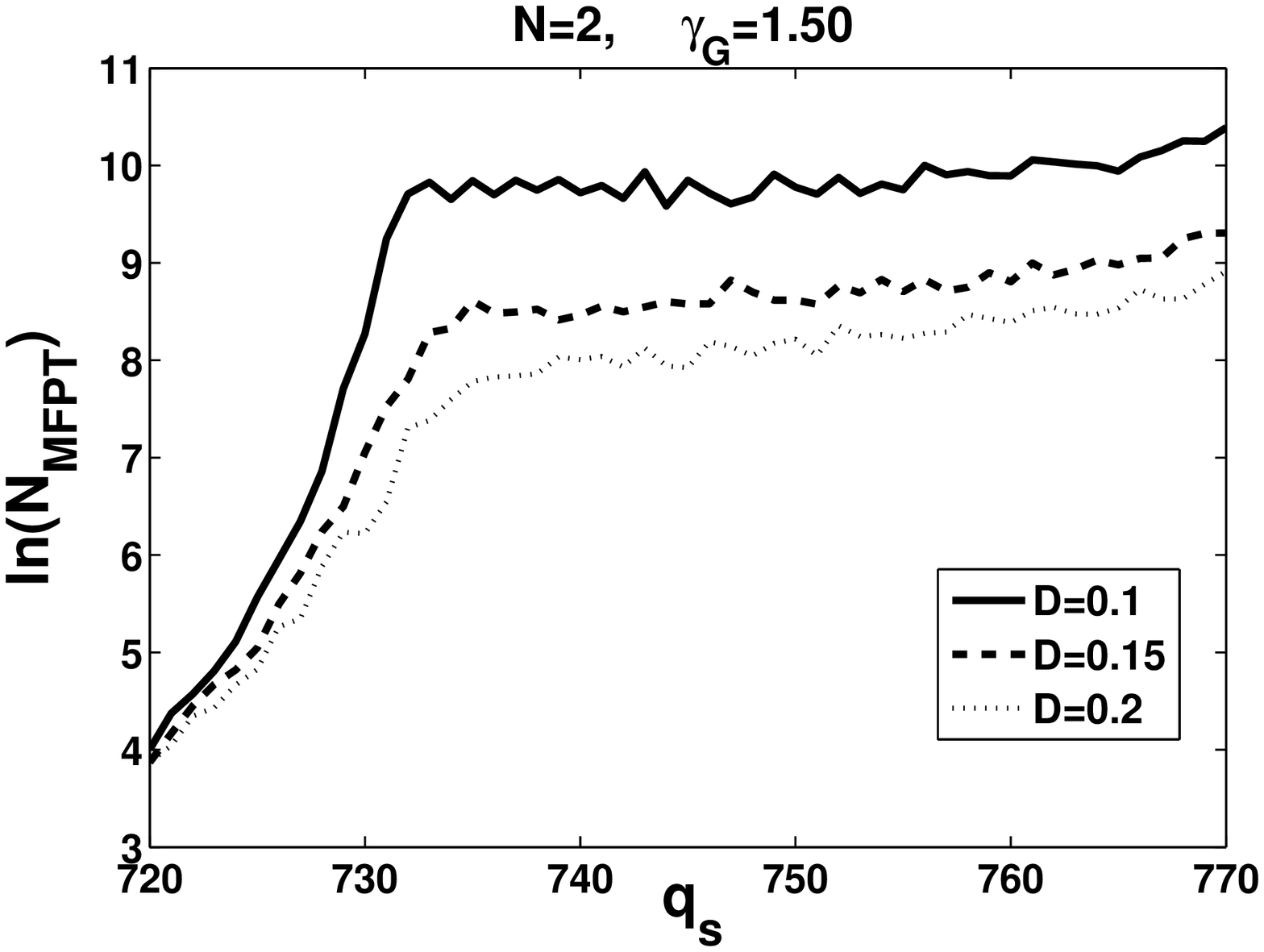}
\includegraphics[scale = 0.35]{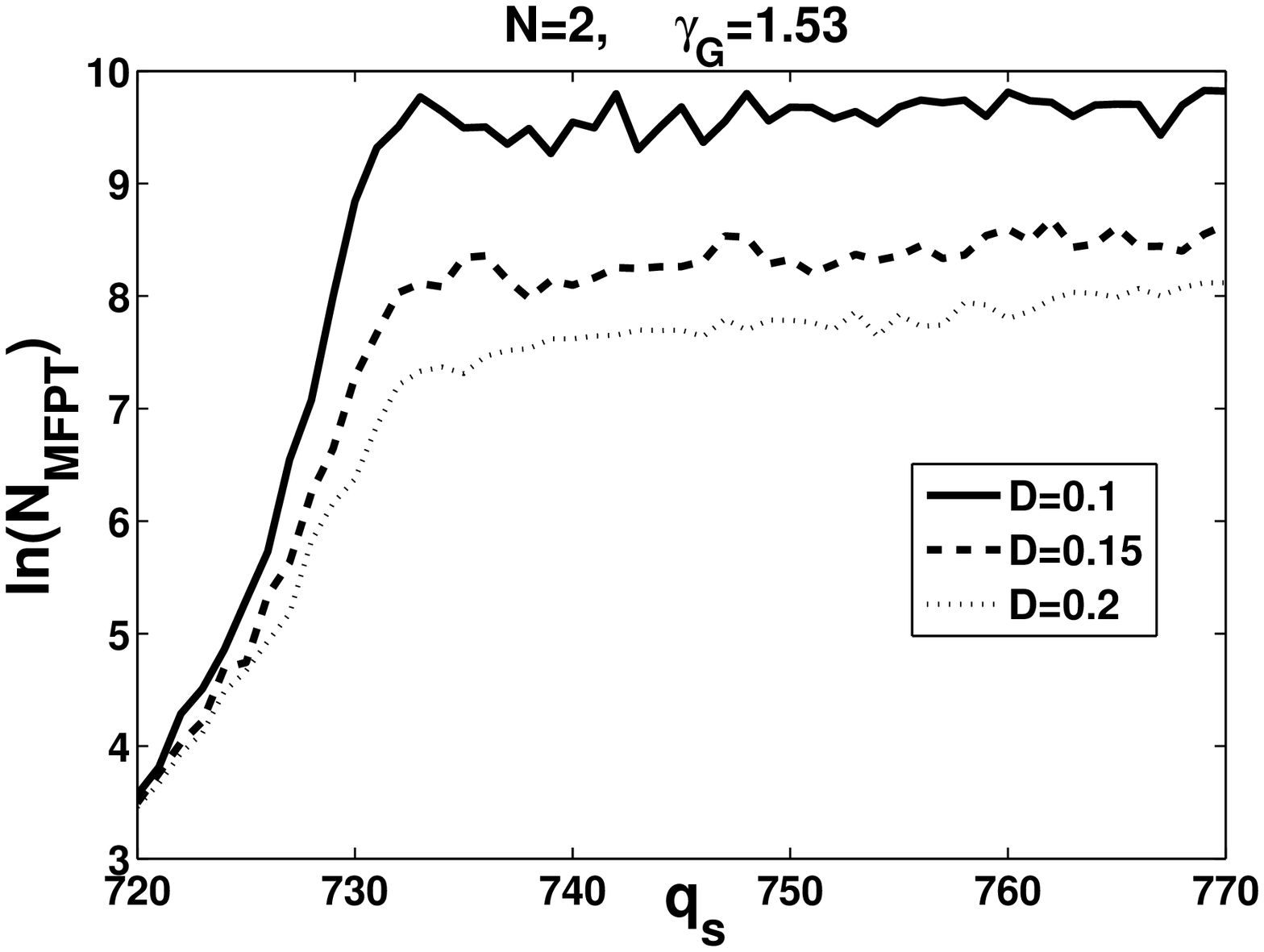}
\includegraphics[scale = 0.35]{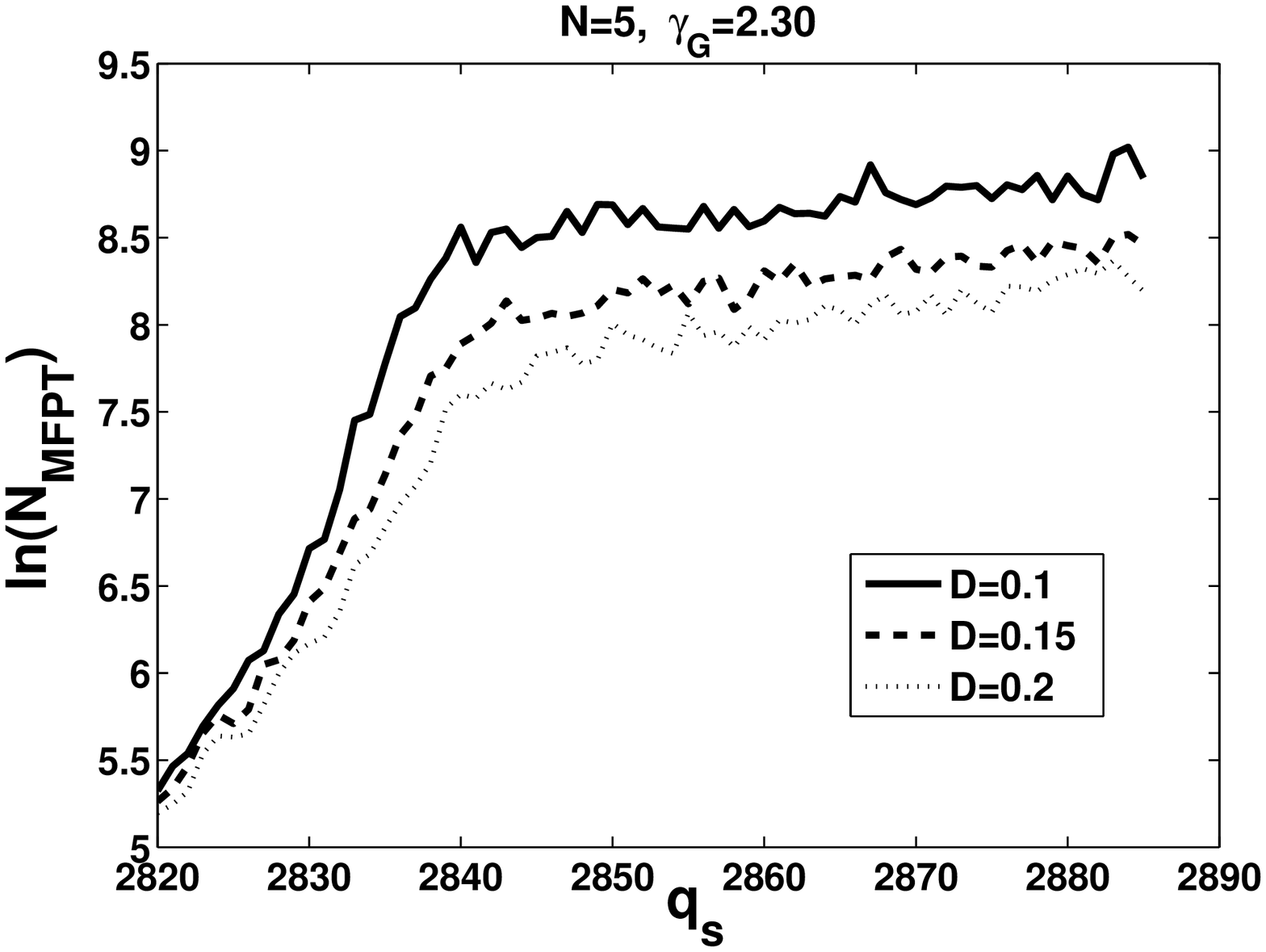}
\includegraphics[scale = 0.35]{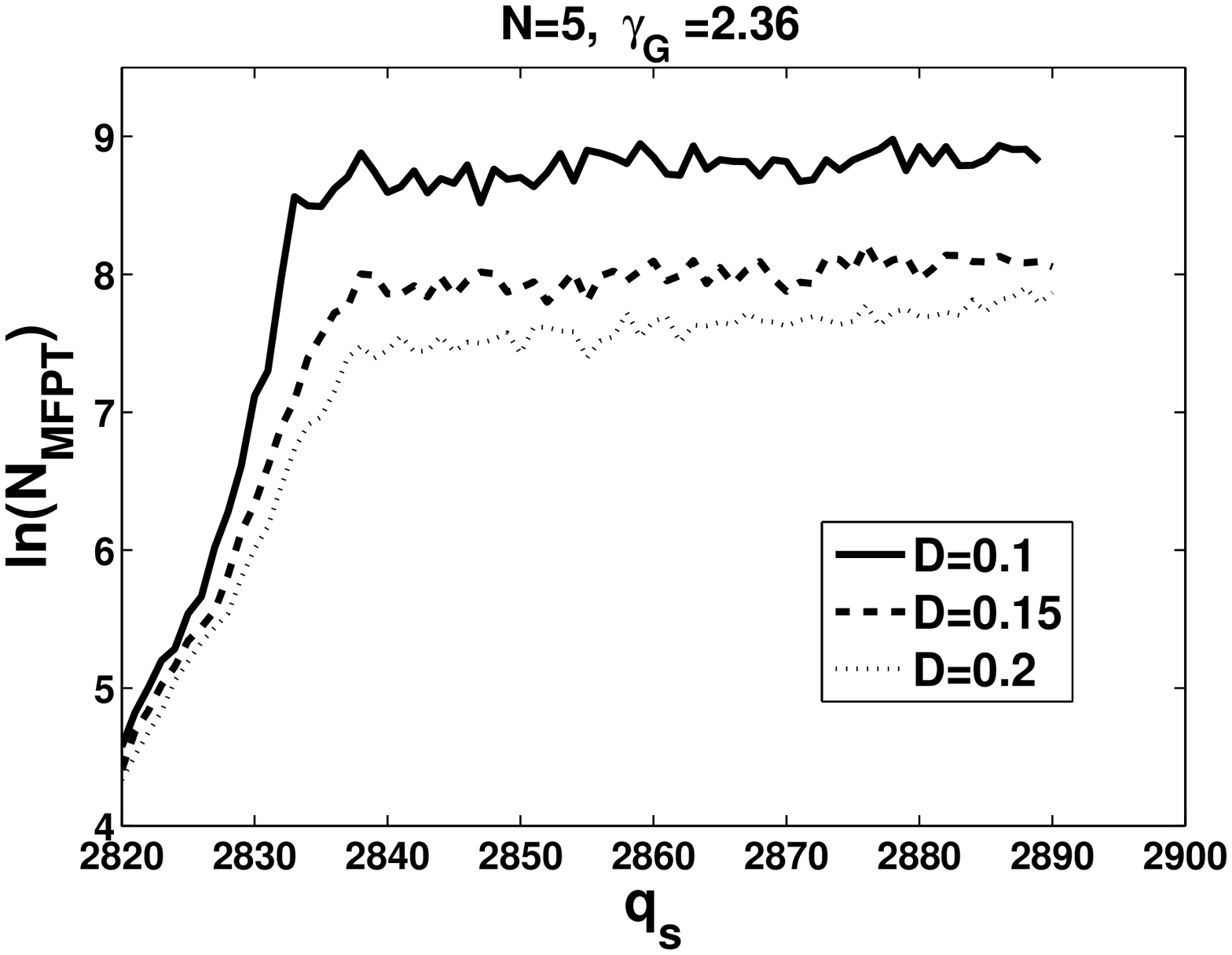}
\caption{\it Average {MFPT} as a function of a threshold $q_s$ at different values of applied current $\gamma_G$ and noise intensity $D$.
It is evident that the method of the knee to identify the separatrix works in a variety of parameters.
The parameters are:
$\beta_L=0.01$, $Q=200$, $\Omega=2.0$, $\alpha=0.1$.
For (i): $N=2$, $\gamma_G= 1.50$; (ii): $N=2$, $\gamma_G=1.53$; (iii): $N=5$,
$\gamma_G=2.30$; (iv): $N=5$, $\gamma_G=2.36$. }
\label{meanfirstp}
\end{center}
\end{figure}

  \begin{figure}
  \begin{center}
\includegraphics[scale = 0.35]{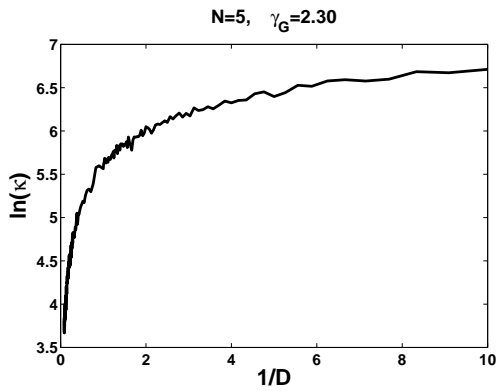}
\caption{\it Example of average of the logarithm of the escape time $\kappa$ from an attractor vs the inverse of the noise  intensity $1 \slash D$.
  The parameters are:
  $\beta_L=0.01$, $Q=200$, $\Omega=2.0$, $\alpha=0.1$, $\gamma_G=2.30$, and $N=5$.}
  \label{escape}
  \end{center}
  \end{figure}

\begin{figure}
\begin{center}
\includegraphics[scale = 0.33]{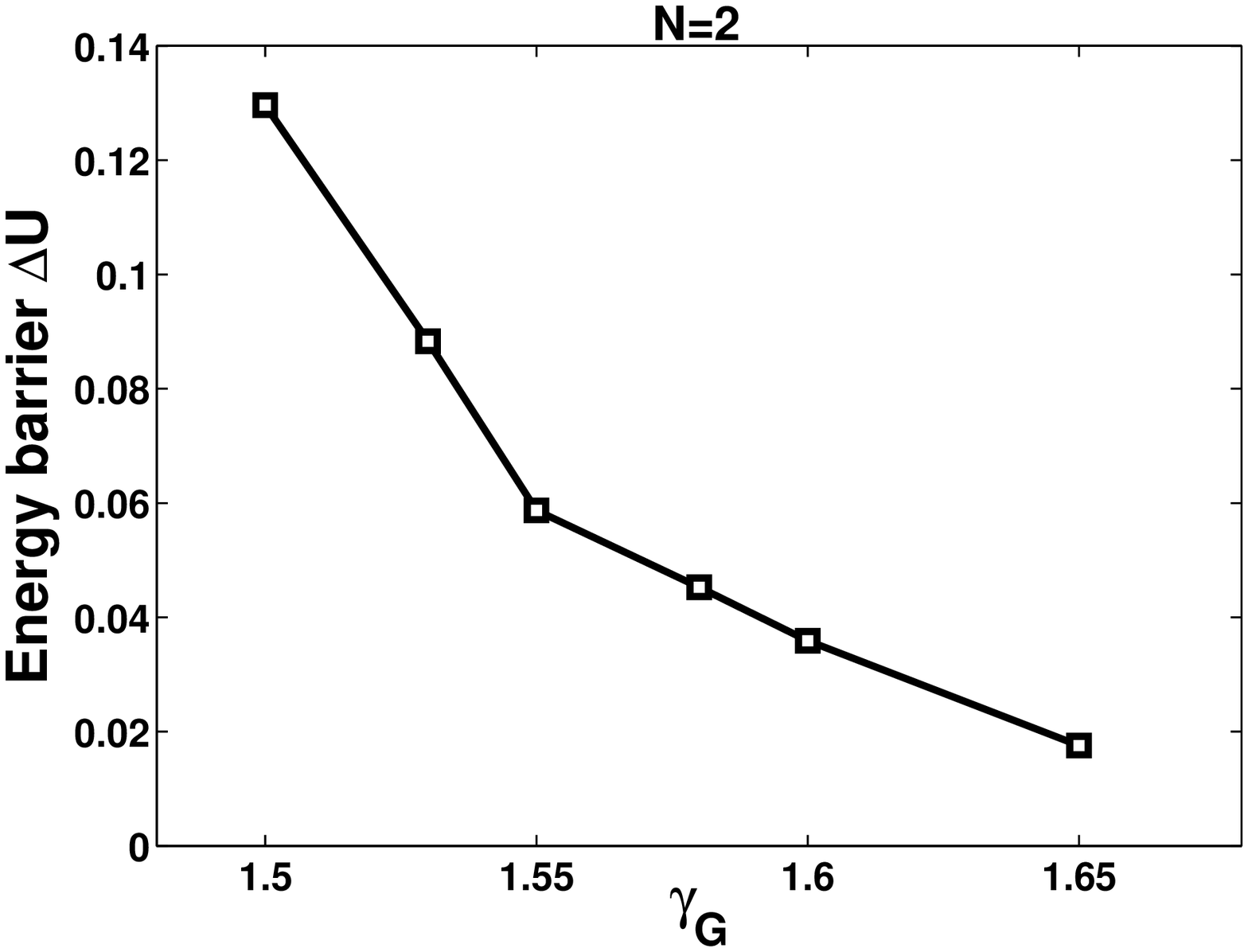}
\includegraphics[scale = 0.33]{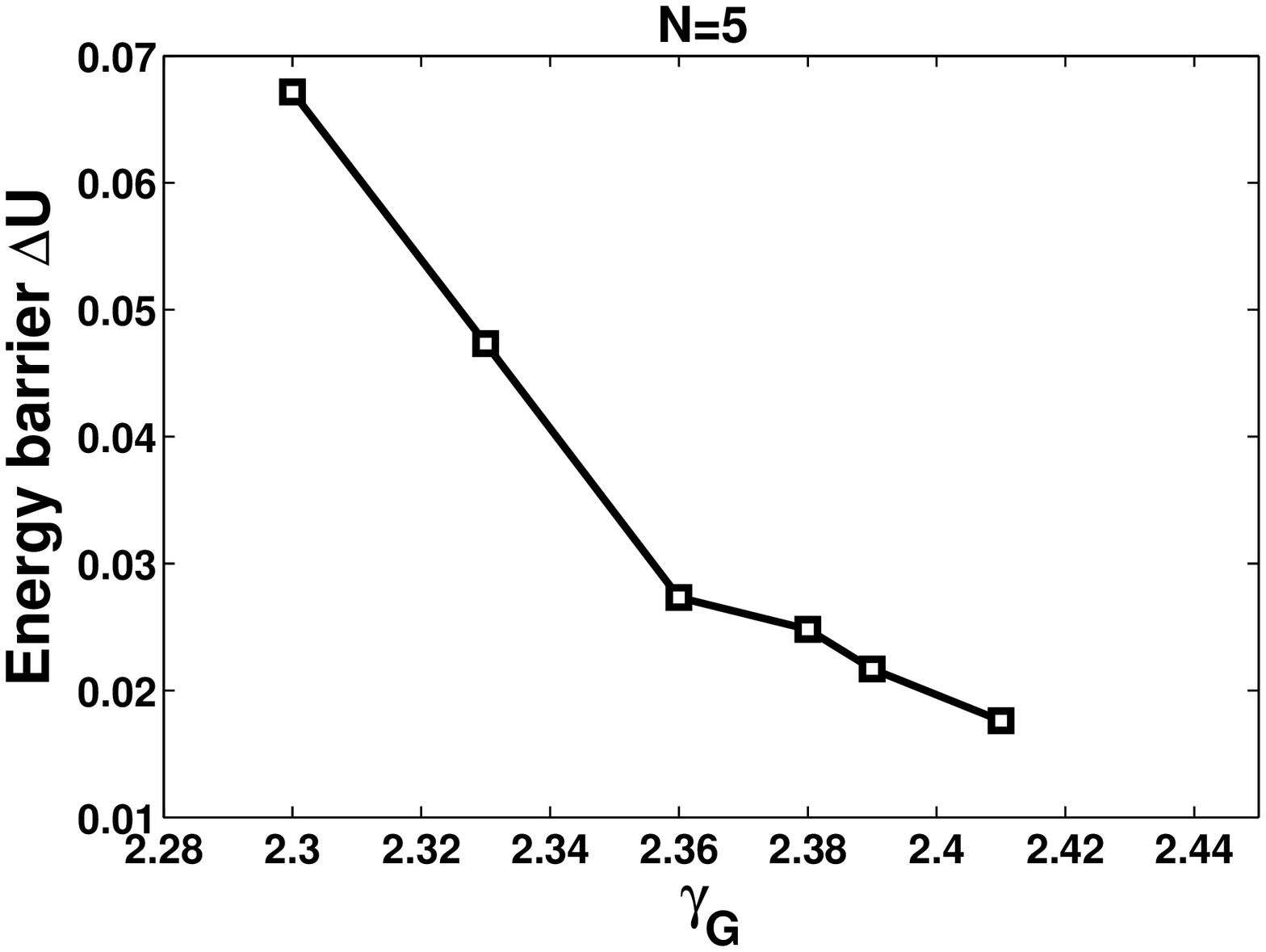}
\caption{\it Energy barrier for the escape from the attractor at frequency $\Omega_1$ (locking to the resonator) as a function of the applied bias current.
The parameters are:
$\beta_L=0.01$, $Q=200$, $\Omega=2$, and $\alpha=0.1$;
 with (i): $N=2$; (ii): $N=5$ .}
\label{barrier}
\end{center}
\end{figure}

\section{Deterministic attractors and stability properties}
\label{deterministic}
{\hg The starting point to retrieve the effects of noise and disorder in the considered array of JJs coupled to a resonator, are the properties of the uniform array.
The dynamics of the ordered array is useful to build a first approximation of the system – e.g. to find the hysteretic IV and the two frequencies (locked and unlocked), as well as the approximate limit cycle and so on. This knowledge is essential to reconstruct the response of the system in the actual disordered and noisy case.
It is important to notice that the information we are aiming to, the effective energy barrier or quasipotential, is defined in the limit of vanishingly noise, and it is therefore conceivable that the properties retrieved in the noiseless case are valid in the limit in which one calculates the quasipotential.
}

\subsection{Attractors properties}
\label{attractors}

In Fig. \ref{I-V} are shown the $IV$ curves for arrays of $2$ and $5$ JJs, obtained increasing and decreasing the bias current $\gamma_G$.
The vertical dashed line denotes a particular bias point the system exhibits two frequencies, $\Omega_1$ and $\Omega_3$ obtained increasing and decreasing the bias current, respectively.
Thus, $\Omega_1$ denotes the frequency on the so-called unperturbed $IV$ curve (some times referred to as McCumber branch) and $\Omega_3$ is related to the resonant frequency of the $RLC$ circuit \cite{Filatrella03}.
{\hg
However, as the main purpose of this work is to illustrate the method of the quasipotential for JJs as birhythmic circuits, we have chosen a set of parameter that best fits this illustration.
In particular, the high value of the coupling strength $1 \slash \beta_L = 100$ tightly couples the JJs parameters to the $RLC$ tanks resonator, that is  therefore strongly influenced by the (nonlinear) inductance  and other JJ parameters (capacitance and resistance).
Thus, the resonance clearly appears to depend upon the number of JJs -- see the shift to higher voltage in Fig. \ref{I-V}(ii).
}
The model is anyway birhythmic, either for the case of a single Josephson junction or an array of JJ: The system exhibits oscillations at two distinct periods depending on the initial conditions.
Table  \ref{Tab1}  summarizes the range of bias current corresponding to the resonant state.
 Figure \ref{attract} displays (i and iii) the projection of the phase portrait space in the $q-dq/d \tau$  plane and (ii and iv) the time evolution of the instantaneous charge.
 The main features are the same as for a single JJ  \cite{Yamapi14}: the branch  locked to the resonator characterized by the frequency $\Omega_1$ exhibits larger excursions of the charge oscillations, while the unlocked branch at the frequency $\Omega_3$ is characterized by smaller oscillations.
\begin{table}
\begin{center}
\begin{tabular}{|c|c|c|}
\hline
$N$ & \multicolumn{2}{c|}{bias currents} \\
\hline
&minimum value& maximum value\\
\hline
$2$ &$ 1.47$ &$ 1.72$  \\
\hline
$5$ &$ 2.28$ &$ 2.42$ \\
\hline
\end{tabular}
\end{center}
\caption{\it Range of the bias current corresponding to JJ locked to the $RLC$ resonant state.}
\label{Tab1}
\end{table}

  \begin{figure}
  \begin{center}
\includegraphics[scale = 0.35]{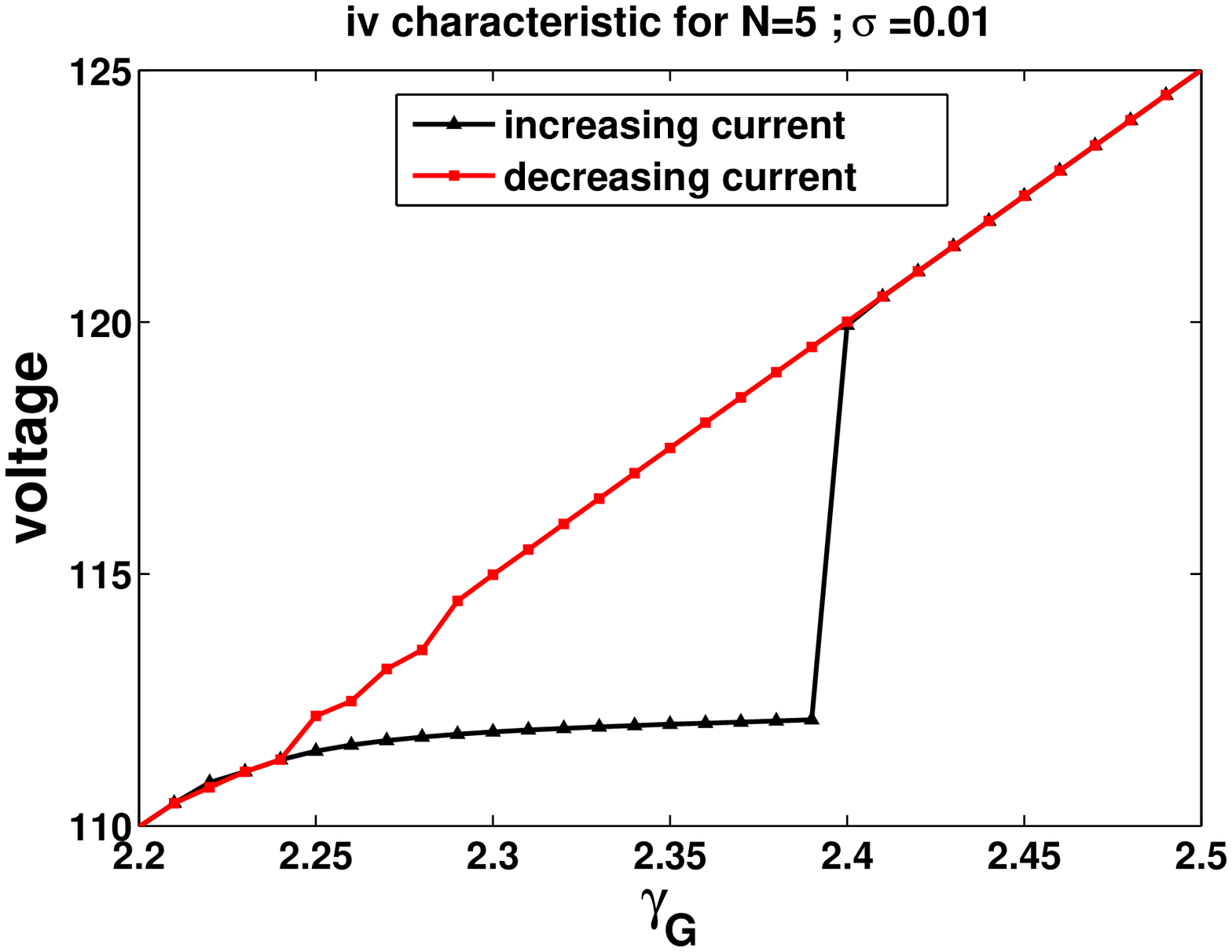}
\includegraphics[scale = 0.35]{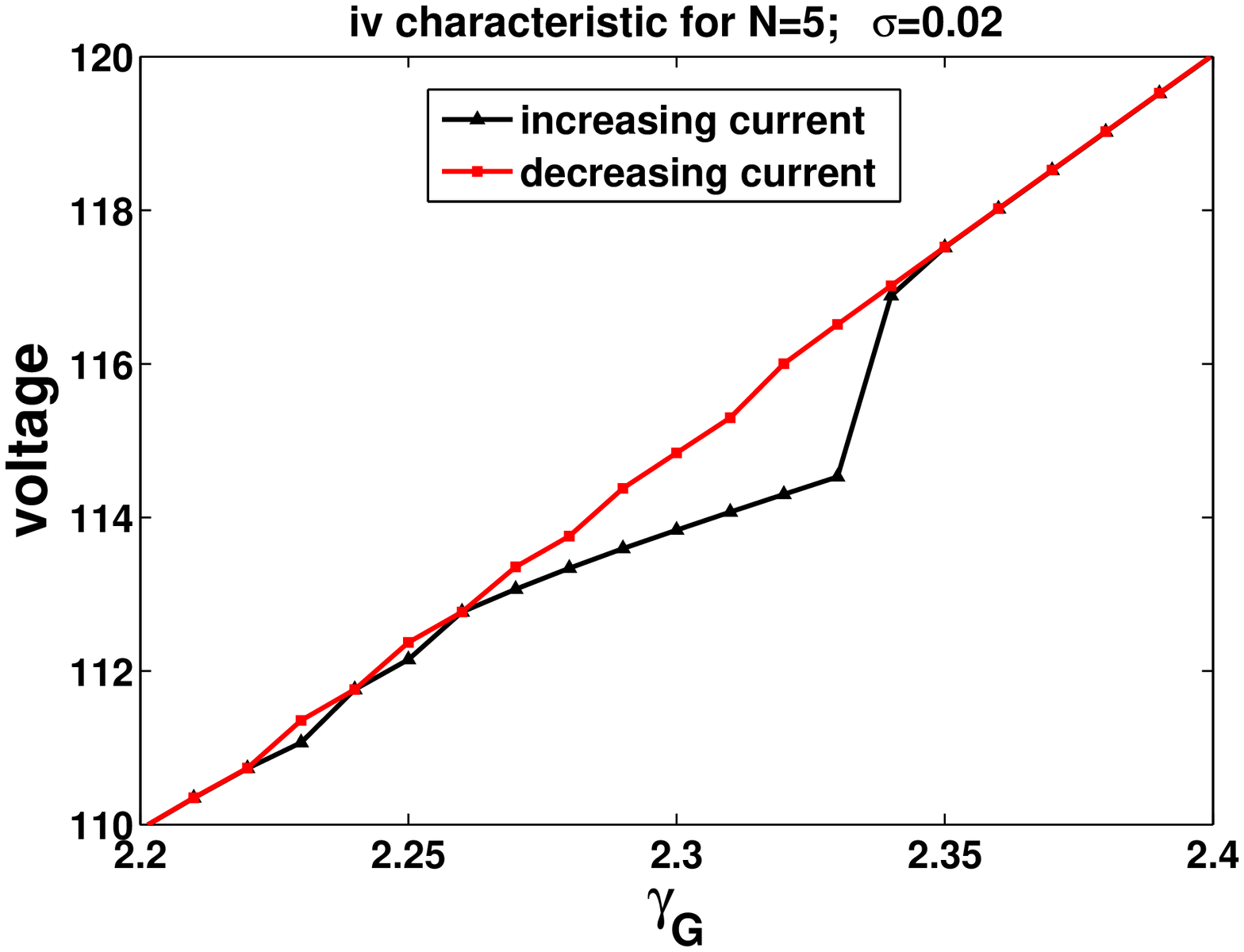}
\includegraphics[scale = 0.35]{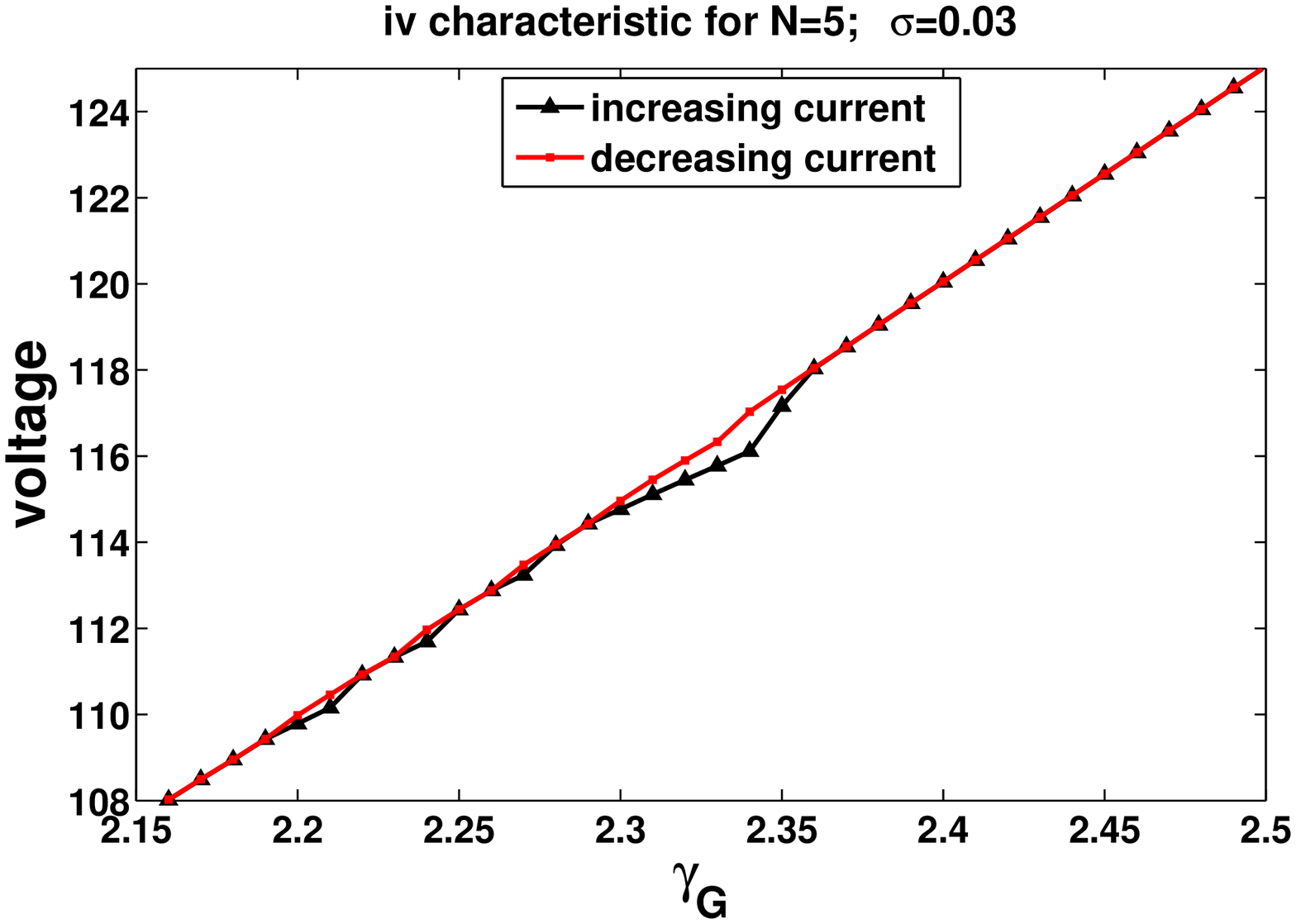}
\includegraphics[scale = 0.35]{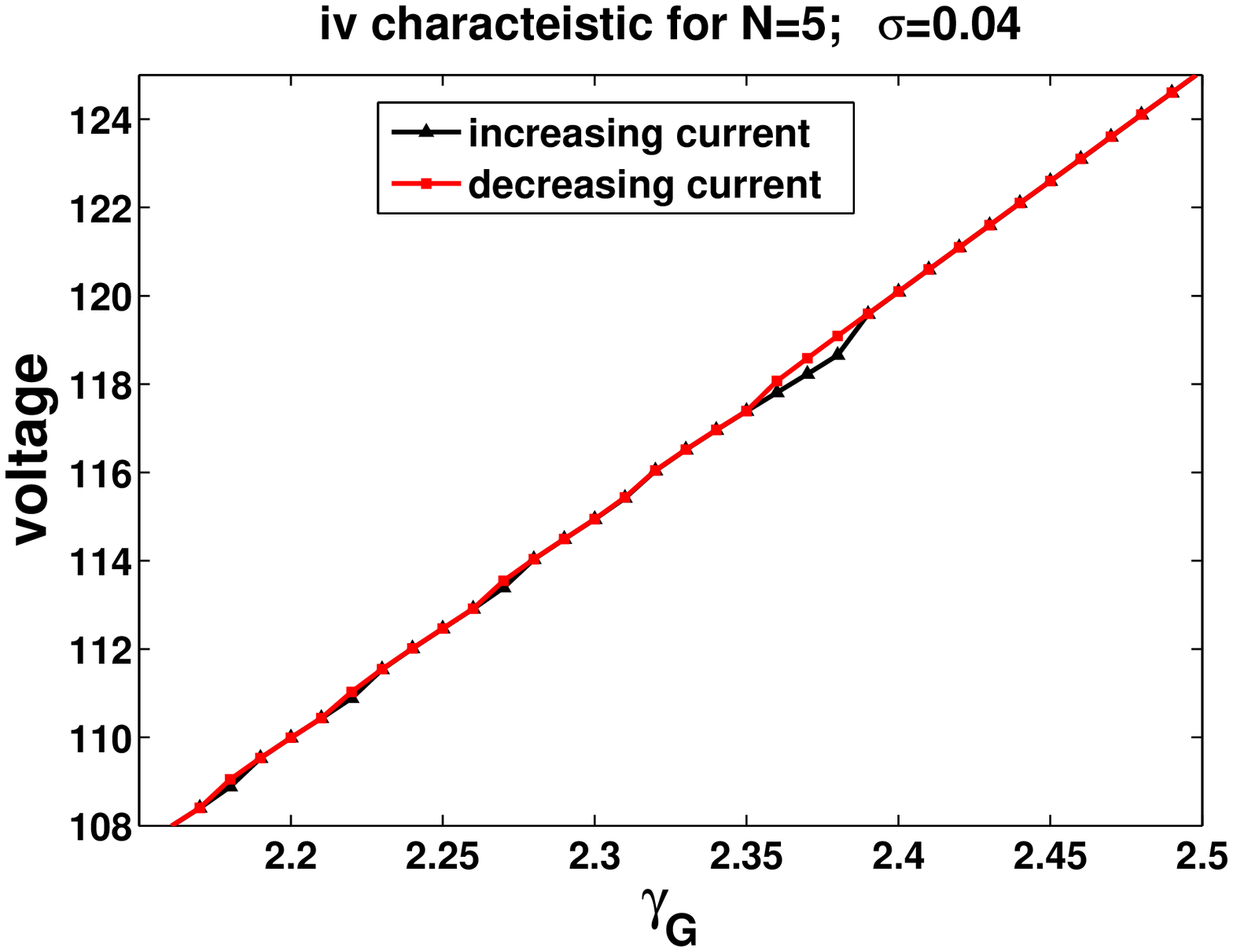}

  \caption{\it Normalized  IV curves of coupled disordered JJ  for   both increasing and decreasing bias current $\gamma_G$.
  The parameters are:
  $\beta_L=0.01$, $Q=200$, $\Omega=2$, $\alpha=0.1$, $N=5$.
(i): $\sigma=1\%$; (ii): $\sigma=2\%$; (iii): $\sigma=3\%$; (iv): $\sigma=4\% $.
}
  \label{disiv}
  \end{center}
  \end{figure}

\begin{figure}
\begin{center}
\includegraphics[scale = 0.35]{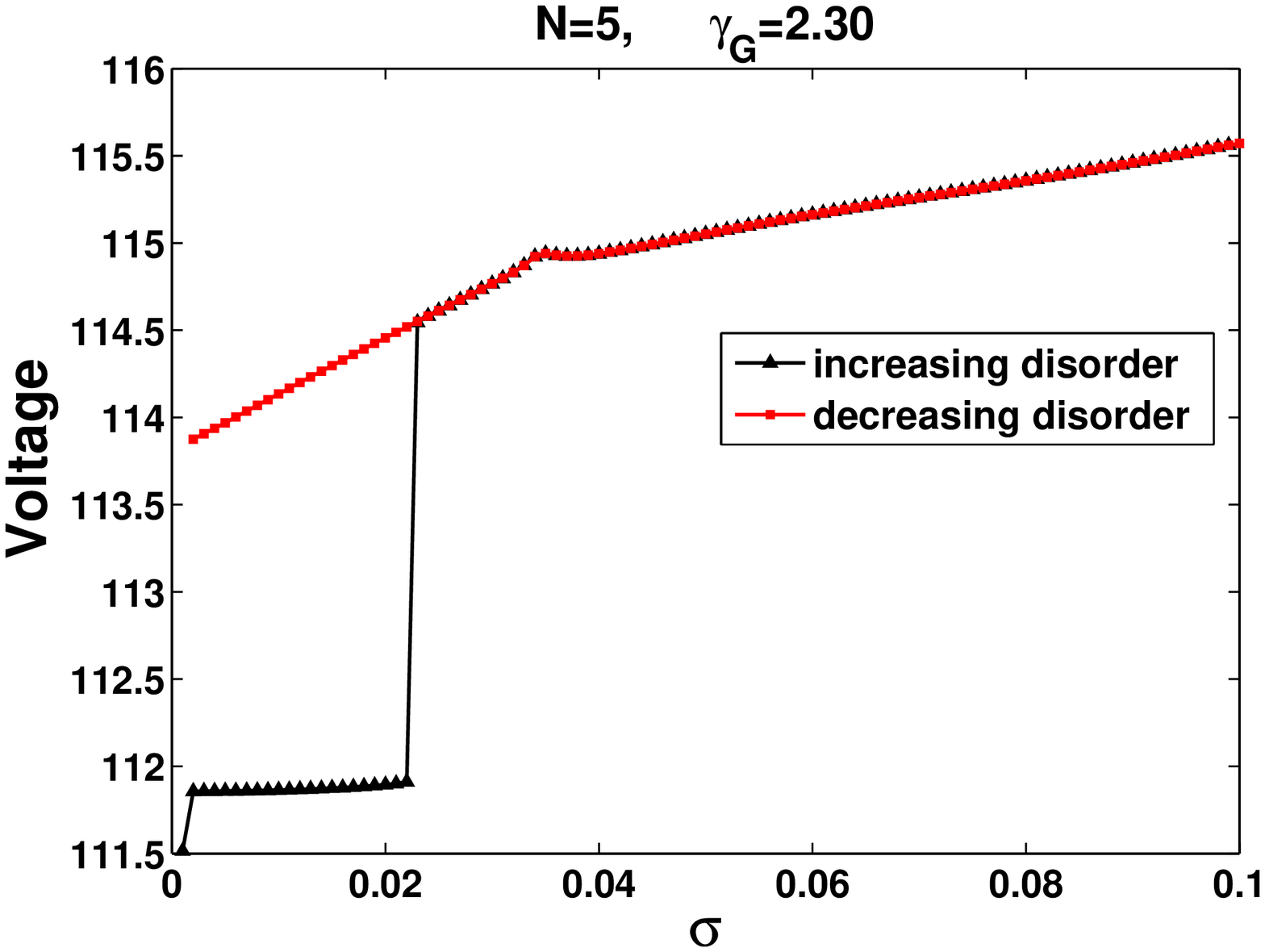}
\includegraphics[scale = 0.35]{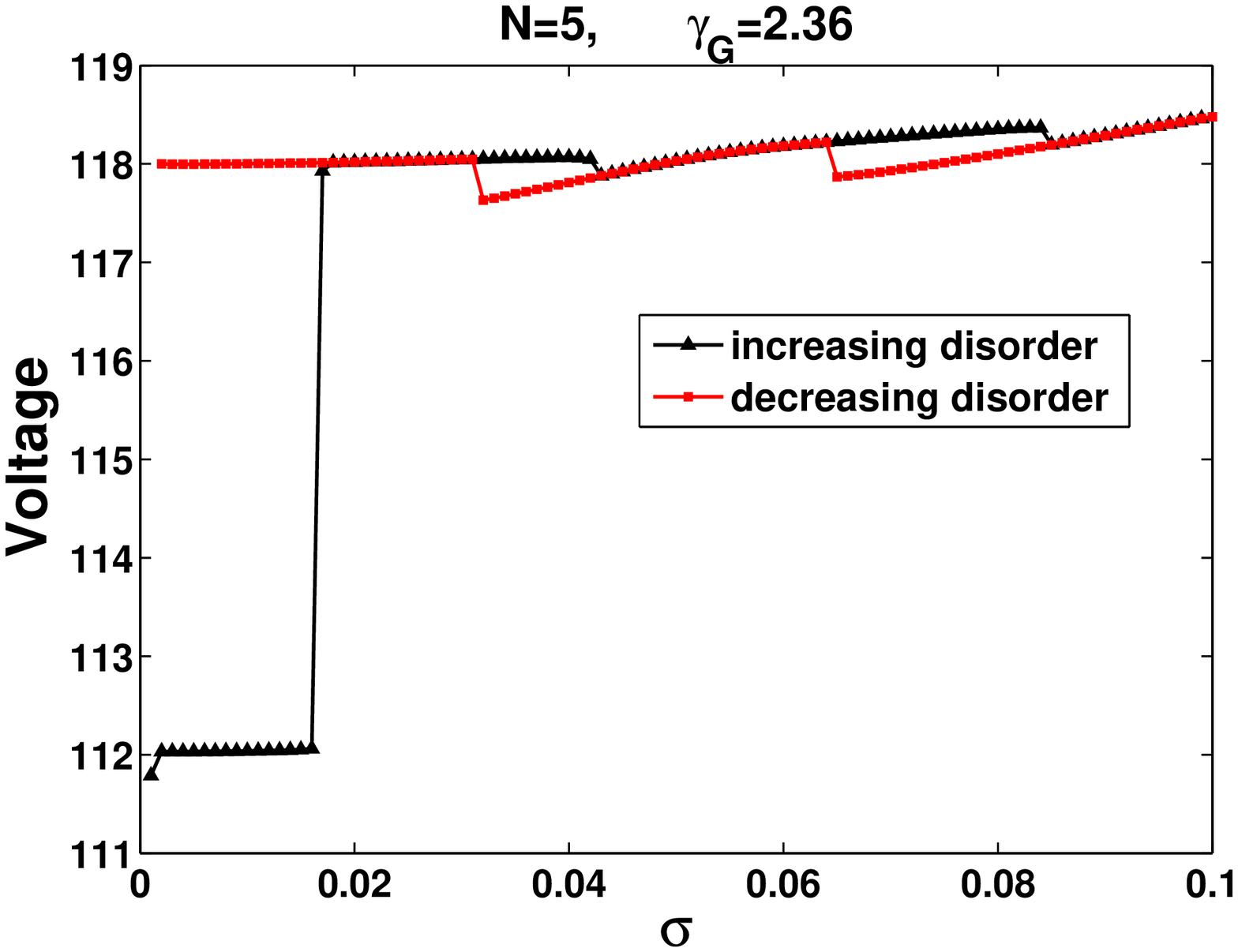}
\caption{\it Normalized voltage as a function of the noise variance $\sigma$.
The parameter are:
 $\beta_L=0.01$, $Q=200$, $\Omega=2.0$, $\alpha=0.1$, $N=5$.
(i): $\gamma_G=2.30$; (ii): $\gamma_G=2.36$ .}
\label{disorder sigmav}
\end{center}
\end{figure}

Between the two stable orbits one postulates the existence of an unstable orbit with frequency $\Omega_2$ that represents the separatrix, whose exact position is not known but can only be estimated \cite{Yamapi14}.
From Figs. \ref{attract} it is evident that as the number of junctions increases the resonant state is moved to larger values of charge, while the amplitude of oscillations remains unchanged.
Thus, the two orbits main features are independent of the number of elements of the model.
 The dynamics under the influence of noise
 shows that the attractors are deformed  but still well separated.
It is therefore possible to estimate the position of the separatrix $q_s$ \cite{Yamapi14}.
The method is summarized in Fig.\ref{switch2}: the examination of the time dependent evolution of the charge $q$ reveals that a sudden switch occurs when the charge suddenly passes from oscillations around a higher value.
During this jump, the charge crosses a threshold  $ q_s$, and for $q > q_s$ the charge increases and then oscillates around a new value (about $q \simeq 780$ in the Figure).
Thus one can roughly estimate the position of the separatrix; however there is clearly an arbitrary in identify  $q_s$.
To refine this guess, it is possible to exploit the properties of the {MFPT}.
In fact the best estimate for the position of the separatrix is signaled by the change of the slope of the {MFPT} $\kappa$ as a function of the threshold $q_s$ \cite{Yamapi14}, as shown in Fig. \ref{meanfirstp}.
The rationale is the following: as the threshold $q_s$ approaches the actual position of the separatrix, the {MFPT} increases exponentially, for the quasi-potential in correspondence of $q_s$ increases.
As the separatrix is passed, the {MFPT} increases much more weakly: beyond the maximum of the quasi-potential the time spent to reach the threshold $q_s$ is but the time to run downhill.
Thus, if the  threshold point $q_s$ is set before the separatrix the {MFPT} increases sharply,  while it weakly increases when the threshold is beyond the separatrix;
 therefore from the change of slope one can estimate the separatrix position.
The knee of the {MFPT}, denoted by the vertical dashed line in Fig. \ref{meanfirstp} is used as an effective separatrix to estimate the energy activation barrier.

The estimate of an energy activation barrier is practically implemented in Fig.\ref{escape} for different value of bias current (and for both $N=2$ and $N=5$).
The linear relationship between the logarithm of the escape time $\kappa$ and the inverse of the noise offers the estimate of an effective  energy barrier $\Delta U$ \cite{Graham86,Kautz88}:
  \begin{equation}
   \label{linearrelation}
   \Delta U = \lim_{D \to 0} D \ln(\kappa).
   \end{equation}
In practical terms, for low enough noise one uses the approximated expression
   \begin{equation}
   \label{logvsinversenoise}
   \Delta U \simeq \frac{\Delta \ln(\kappa)}{\Delta(1 \slash D) }.
   \end{equation}
Equation (\ref{logvsinversenoise}) is very important to characterize with an activation energy the metastable state in the birhythmic region, that is the subject of next Section.

  \subsection{Energy barriers}
\label{energy}

\begin{figure}
\begin{center}
\includegraphics[scale = 0.3]{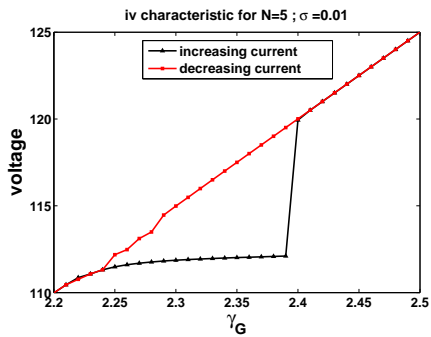}
\caption{\it Normalized  IV curves for the disordered coupled {JJ} model for both
increasing and decreasing bias current $\gamma_G$.
The parameters are:
$\beta_L=0.01$, $Q=200$, $\Omega=2.0$, $\alpha=0.1$, $N=5$, $ \sigma=0.1\%$. }
\label{disorderiv0001}
\end{center}
\end{figure}

\begin{figure}
\begin{center}
\includegraphics[scale = 0.35]{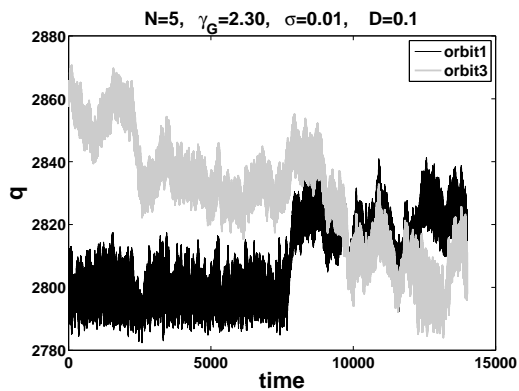}
\caption{\it Example of a switch from the locked attractor at the $RLC$ frequency $\Omega_1$  to the unlocked at the McCumber frequency $\Omega_3$ under the influence of noise for the disordered JJ.
Between the initial time and the normalized time $5000$ system crosses the separatrix $q_s$.
The parameters are:
$\beta_L=0.01$, $Q=200$, $\Omega=2$, $\alpha=0.1$, $\gamma_G=2.30$, $\sigma=0.1\%$, $D=0.2$, $N=5$.}
\label{sig0001switch5}
\end{center}
\end{figure}

 In this Section, we analyzed the behavior of the activation energy as a function of the bias point, see Fig. \ref{I-V}.
 In the two panels of Fig.\ref{barrier}  the common feature is that the  activation energy is low ($0.02\leq \Delta U \leq 0.14$ for $(i)$  and $0.02 \leq \Delta U \leq 0.07$ for $(ii)$).
 At the bottom of the step the energy barrier is at a maximum; while the current is increased  along the step the energy barrier decreases and almost disappears   at the top of the step, for the gap energy decreases when the number of lumped elements increases.
    Thus the system is less stable as the number of elements increases.
    According to the lifetime of the $RLC$-induced step, it decreases along the step when the current bias $\gamma_G$ increases independently of the number of the lumped elements, see Fig.\ref{escape}.

  \section{Effects of disorder in the model}
\label{disorder}
JJs are fabricated with photo-lithographic processes, and are therefore each JJ is different from the other \cite{Tolpygo15,Tolpygo17}.
It is relevant to investigate how the differences in the parameters reflect on the synchronization properties.
The effect of  disorder, respect to ideal arrays of identical JJs  is the subject of this Section.

  \subsection{IV Characteristics}

To retrieve the IV characteristics of noiseless arrays, Eqs. (\ref{eqJJ_RLC_disordered}) and (\ref{epsuniform}) are  simulated without noise.
Figs \ref{disiv} show the resulting IV for increasing values of the disorder parameter $\sigma$.
From the data it is evident that as the disorder parameter increases, birhythmicity disappears, and the system remains  birhythmic only for the low values of the disorder parameter, $\sigma < 4\%$.
The behavior is confirmed by the diagram of the voltage as a function of disorder in Fig.\ref{disorder sigmav}.
The curves are obtained starting from the $RLC$ locked state for uniform JJ, and then slowly increasing the disorder up to $\sigma = 10\%$.
The procedure is then reversed, and the disorder is slowly decreased.
In Fig. \ref{disorder sigmav}(i) it is evident that disorder induces de-synchronization.
This is complementary to the observation that varying the number of junctions the arrays lock to the cavity \cite{Filatrella03}.

\subsection{Attractors properties}

As the birythmicity of the system remains for low values of the disorder $\sigma$, simulations have been performed to investigate the attractors for low disorder values (e.g., $\sigma= 0.1\%, 0.3\%, 0.55 \% ...$).
The properties are very similar, therefore one can focus only  on $\sigma=0.1\%$.
The resulting IV curve is displayed in Fig. \ref{disorderiv0001}, that is very similar to the uniform case of Fig. \ref{I-V}(ii).
Following the behavior display on Figure 5, one finds that
for the low disorder case the amplitude of the oscillations at the frequency $\Omega_1$ (the $RLC$ frequency) is much larger than the oscillations at the frequency $\Omega_3$ (the unlocked or McCumber solution \cite{Barone82}).
At variance with the uniform case of Fig. \ref{attract}, the oscillations are much smaller.
This is the first effect of the disorder: the unlocked oscillations of the charge almost disappear.
 With $D\neq 0$, it is evident that the attractors are still clearly separated.
 However, the transition from an attractor to the other is much smoother,
  as shown in Fig. \ref{sig0001switch5}.
 Under the combined effect of noise and disorder the passage
 from an attractor to another is much less sharp, and it
 is therefore much more difficult to identify the separatrix $q_s$.
This is the second remarkable effect of disorder: the switch from the orbit of frequency $\Omega_3$ to the orbit at frequency $\Omega_1$ is less neat, and therefore the position of the separatrix cannot be retrieved using the mean first  passage time method of Fig. \ref{meanfirstp}.
This change induced by disorder entails the  difficulty to evaluate the energy barrier and the study of  the global stability.

\section{Conclusions}
\label{conclusions}
{\hg This work analyzes the behavior of coherent cooperation of Josehson junctions, a topic of interest for practical reasons (e.g., increase the emitted power \cite{Tachiki11,Welp13}) and as prototype for synchronization.
Series arrays of identical {Josephson Junction} coupled through a linear $RLC$ resonator  behave qualitatively as a single junction, for instance the system exhibits two clearly distinct frequencies in the locked and unlocked (to the resonator) cases.
The approximation of identical {Josephson Junction} is unrealistic, for the fabrication process produces changes from  junction to junction.
When disorder and noise are included, some special features of the arrays emerge, in particular the possibility of large excursions that drive the system from the locked to the unlocked state.
Such large excursions cannot be treated in the framework of the
 Kuramoto model, that deals with the local stability properties \cite{Rodrigues16}.
An alternative approach based on the quasipotential method \cite{Graham86,Bouchet16}
has proved fruitful for the single junction case, for both the voltage standard application in which the JJ is driven by an external rf source  \cite{Kautz96} and of a single JJ coupled to a resonator \cite{Yamapi14}.
We have extended the application of the method to a series array coupled to an $RLC$ resonator.
To make the extension possible it has been necessary to identify the effective separatrix -- the passage from the locked to the unlocked phase space region.
This effective border is difficult to determine, even in the noiseless and ordered case, for the system is high dimensional ($2N+2$ dimension for $N$ junctions).
The employed method is an approximate one: supposing that the separation region is just a plane identified by the a single coordinate (the charge on the resonator), it is possible to compute the MFPT to cross this border as a function of the charge threshold, in analogy with the single JJ case, for the MFPT behavior suddenly changes when the threshold is passed.
Numerical findings are encouraging: the change in the slope is neat, and can be clearly identified also in the case of multiple JJ, up to $N=5$.
Thus, it is possible to compute the effective confining energy for ordered multiple JJ coupled to a resonator.
The numerical investigation has revealed two important features.
First, the behavior of this effective energy is similar to the case of a single JJ: it is higher at the top of the resonant step, and gets smaller towards the bottom.
Second, the effective energy barrier decreases with the number of JJ, making it easier a passage from the locked to the unlocked state when more JJs are present.
In the presence of disorder the situation looks more complicated: as the locked and unlocked states get closer to each other, and therefore it is more difficult to identify, with the above method, the effective separatrix.
However, for very low disorder variance the method is applicable and it appears that the system retains the qualitative features of the passages from one state to the other.
Some limits of the present study are evident. First, this is a proof of principle for relatively few JJ and secondly to a specific configuration of the external load.
It would be interesting to extend the investigation to other configurations \cite{Shukrinov12,Shukrinov17}, to investigate the role of the resonator and JJ parameters,  and to consider many more JJs. The latter  case calls for much more demanding numerical simulations, that presumably are only possible with massive parallel computations, possibly on cheap CUDA hardware \cite{Pierro18}.
}
\section*{ Acknowledgements }

The authors thank V. Pierro for useful discussions.

\end{document}